\documentclass[%
 aip,
 amsmath,amssymb,floatfix,
reprint,
]{revtex4-1}

\usepackage{graphicx}% Include figure files
\usepackage{dcolumn}% Align table columns on decimal point
\usepackage{bm}% bold math
\usepackage[utf8]{inputenc}
\usepackage[T1]{fontenc}
\usepackage{mathptmx}
\usepackage{etoolbox}
\usepackage{appendix}

\newcommand{\bnabla}{\mbox{\boldsymbol{$\nabla$}}}

\makeatletter
\def\@email#1#2{%
 \endgroup
 \patchcmd{\titleblock@produce}
  {\frontmatter@RRAPformat}
  {\frontmatter@RRAPformat{\produce@RRAP{*#1\href{mailto:#2}{#2}}}\frontmatter@RRAPformat}
  {}{}
}%
\makeatother
\begin{document}

%\preprint{AIP/123-QED}

\title[3D Magnetic Separator Reconnection]{On the importance of separators as sites of 3D magnetic reconnection}
\author{C.E. Parnell}
 \email{cep@st-andrews.ac.uk}
\affiliation{School of Mathematics and Statistics, University of St Andrews, St Andrews, Fife, KY16 9SS, Scotland, UK}

\date{\today}

\begin{abstract}
For 3D magnetic reconnection to occur there must exist a volume within which the electric field component parallel to the magnetic field is non-zero. In numerical experiments, locations of non-zero parallel electric field indicate sites of 3D magnetic reconnection. If these experiments contain all types of topological feature (null points, separatrix surfaces, spines and separators) then comparing topological features with the reconnection sites reveals that all the reconnection sites are threaded by separators with the local maxima/minima of the integrated parallel electric along fieldlines coinciding with these separators. However, not all separators thread a reconnection site.

 Furthermore, there are different types of separator. Cluster separators are short arising within an individual weak magnetic field region and have little parallel electric field along them so are not associated with much reconnection. Intercluster separators connect a positive null point lying in one weak-field region to a negative null point that lies in a different weak-field region. Intercluster separators often thread enhanced regions of parallel electric field and are long. Since separators form the boundary between 4 globally-significant topologically-distinct domains, they are important sites of reconnection which can result in the global restructuring of the magnetic field. 

By considering kinematic bifurcation models in which separators form it is possible to understand the formation of cluster and intercluster separators and explain their key properties.
\end{abstract}

\maketitle

\section{Introduction \label{sec:introduction}}
Magnetic reconnection was first proposed around 65 years ago\citep{Parker1957,Dungey1961}. Despite many advances,
especially in understanding 2D reconnection, key aspects of this fundamental plasma
physics process still remain poorly understood in 3D.

It is well known that magnetic reconnection in two-dimensions (2D) occurs at magnetic null points where $\boldsymbol{B}=\boldsymbol{0}$, 
for example, [\onlinecite{MagRecon00}]. 
In 2D such points are either elliptic (O-type) or hyperbolic (X-type). About X-type nulls, there lie 4 topologically distinct flux regions separated by 4 special fieldlines (separatrices) that extend to/from the null point and from the null's `X' structure.
X-type null point reconnection involves the transfer of magnetic flux between one pair of oppositely-situated topologically-distinct flux regions into the other pair. 
Specifically, a stagnation-type flow is associated with 2D reconnection at an `X' point, with the inflow sweeping magnetic field in towards the null where reconnection causes a ``cutting'' of a pair of fieldlines creating 4 fieldline segments that are rejoined such that they create a new pair of fieldlines. 

About an O-type null, there is just the one topologically distinct region, and so flux is either created or destroyed at the null point.

As revealed by Schindler and Hesse\citep{Schindler88, Hesse88}, in three dimensions (3D) magnetic reconnection occurs at locations where the component of the electric field parallel to the magnetic field (from now on called the parallel electric field) is non-zero. Despite the fact that in 3D  
this is a necessary requirement for reconnection, many researchers do not consider the parallel electric field when attempting to identify reconnection sites nor when analysing the magnetic reconnection taking place in the experiment. Unfortunately, this means that erroneous conclusions may be reached about the location and nature of the reconnection. 
This is because magnetic reconnection requires the correct magnetic geometry or topology and also the correct plasma flow/conditions to produce an appropriate electric field. 
\begin{figure}
  \centering
  \includegraphics[scale=0.5]{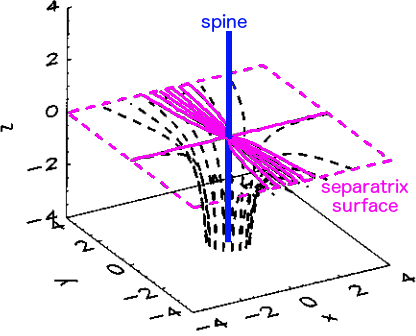}
  \includegraphics[scale=0.55]{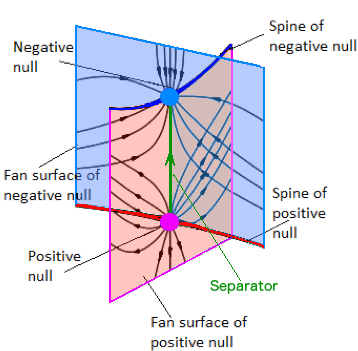}
  \caption{Plots showing the structure and fieldlines of (top) a
  3D improper null and (bottom) a positive null point and a negative null point connected by a separator.} 
\label{fig:posnull-sep}
\end{figure}
As in 2D, reconnection may also occur in the vicinity of 3D null points (see for example [\onlinecite{PontinPriest2009}]).

The fieldlines that extend out from or into a 3D null point form a separatrix surface and a pair of spines (Figure~\ref{fig:posnull-sep}, top)\citep{Fukao1975,Greene1988,Lau1990,Parnell1996}. If the fieldlines forming the two spines head into the null, then the lines lying in the separatrix surface will all be directed outwards from the null. Such a null is called a positive null. If the separatrix surface fieldlines are directed in towards the null and the spines outwards, then the null is called a negative null. 
Alternative names have been given to these 3D nulls such as A-type and B-type for negative and positive nulls, respectively\citep{Greene1988,Lau1990} or the nulls have been identified according to their topological degree: positive (negative) nulls have a topological degree of $-1$($1$), somewhat confusingly\cite{Fukao1975,Deimling1985,Greene1988}.

There exist special fieldlines that start from a positive null and end at a negative null and so lie within the separatrix surfaces of two null points (Figure~\ref{fig:posnull-sep}, bottom). Such fieldlines are called separators and it may be argued that these 3D structures are the topological equivalent in 3D to a 2D null. This is because both of these structures form the boundaries between four topologically distinct flux domains %\textbf{(see later[be specific] for more details)}. 
Therefore, reconnection at separators may permit significant transfer of flux between different flux domains and can also lead to the creation of new topologically distinct flux domains (see for example [\onlinecite{Haynes2007,Parnell2008,
Wilmot-Smith2011,Stevenson2015a}]). Hence, just like 2D reconnection at X-points, separator reconnection can produce global restructuring of the magnetic field.

Note, that the intersection of two separatrix surfaces are generic, stable separators\cite{Lau1990,BrownPriest1999}. Other separators formed, for example, by a spine line from a negative null point ending at a positive null point (e.g., spine lines from two null points coinciding) are unstable and will be destroyed by any perturbation of the magnetic field. The same is true for a separator temporarily formed when a spine line from one null lies in the separatrix surface of another null of the same sign\cite{BrownPriest1999}. These two types of separator are non-generic separators and are always short-lived. They do not form a boundary between topologically distinct domains and are unlikely to be sites of reconnection.  

Separators have been considered in a number of papers. Lau \& Finn\cite{Lau1990} presented a model for a separator ring in which a pair of separators (which they called X-lines) connecting an opposite polarity pair of null points. Their separators are formed by the intersection of the separatrix (which they called $\sum$-) surfaces from the two nulls. The spine (which they called $\gamma$) lines provide a bound to two edges of the separatrix surfaces. The resulting structure looks like the intersection of a pair of cones, one on its base the other on its point. They also considered a 3D kinematic reconnection model explaining the transfer of flux across the separator. Their analysis identified that about the separator the field lines form an $X$-like structure if the component of magnetic field parallel to the separator is ignored. Longcope \& Cowley\cite{Longcope1996} look at the formation of a current sheet at such a separator ring.

Haynes et al.\cite{Haynes2007} noted during the analysis of a numerical experiment involving separators that the fastest reconnection across a separator occurs when the magnetic field lines are twisted about the separator and thus form an $0$-type structure if the component of magnetic field parallel to the separator is ignored. The latter implies that the fastest reconnection occurs where the electric current density parallel to the separator is largest, which is not unreasonable. 

Brown \& Priest\cite{BrownPriest1999} discuss a range of topologies involving separators, but their analytical models mainly consider point sources on the base of the domain. They identify two separator bifurcations which they call the local or global separator bifurcations which are discussed later.

Various models of both the solar atmosphere (e.g., [\onlinecite{Platten2014}]) and also the Earths magnetosphere (e.g. [\onlinecite{Dorelli2007,Komar2013}]) have been developed in which multiple separators have been identified. Since separators are the topological structures bounding four topologically distinct flux domains they are important locations about which reconnection can lead to the global restructuring of the magnetic field.

The main focus of this paper is reconnection at generic separators. Below (Section~\ref{sec:reconnection-equations}) we present a brief explanation on why parallel electric fields, however they are generated, are required for 3D magnetic reconnection. Then a brief review of papers that have compared topological features with sites of reconnection identified by calculating parallel electric fields is presented in Section~\ref{sec:num-experiments}. In Sections~\ref{sec:creation of nulls} and~\ref{sec:separators}, the magnetic bifurcations that lead to the creation of separators are presented. The importance of separators, formed during the bifurcations, as sites of reconnection are discussed in Section~\ref{sec:separator reconnection}. Then, Section~\ref{sec:sep-form-mechanism} discusses how these bifurcations result in the creation of two distinct types of separator. Finally, the conclusions are presented in Section~\ref{sec:conclusions}.

\section{Three-dimensional Magnetic Reconnection \label{sec:reconnection-equations}}

\subsection{Magnetic Induction Equation \label{sec:induction-equation}}
The equation governing the time evolution of the magnetic field comes from combining Faraday's law, 
\begin{equation}
\label{eqn:faradays}
    \frac{\partial \boldsymbol{B}}{\partial t} = -\bnabla\times\boldsymbol{E}\;,
\end{equation}
where $\boldsymbol{B}$, $\boldsymbol{E}$ and $t$ represent magnetic field,  electric field and time, respectively, and Ohm's law, 
\begin{equation}
\label{eqn:gen-ohms}
\boldsymbol{E}+\boldsymbol{v}\times \boldsymbol{B} = \boldsymbol{N}\;,
\end{equation}
where $\boldsymbol{v}$ is the plasma velocity and $\boldsymbol{N}$ represents all non-ideal terms. By eliminating $\boldsymbol{E}$ the Induction equation is formed and is given by
\begin{eqnarray} 
\label{eqn:gen-induction}
\frac{\partial \boldsymbol{B}}{\partial t} =  \bnabla \times \left(\boldsymbol{v}\times \boldsymbol{B}\right) - \bnabla \times \boldsymbol{N}\;.
\end{eqnarray}
The first term on the right-hand side of this equation is the advection term and the second is the diffusion term. 
The non-ideal terms, $\boldsymbol{N}$, involved in a Generalised Ohm's law may include not only terms relating to electric current densities but also electron and ion pressures and inertias, as well as the Hall term. The ratio of the advection term over the diffusion term forms the dimensionless magnetic Reynolds number, $R_m$. In the perfectly conducting limit, the magnetic Reynolds number is very large and the diffusion term can be ignored. Under these conditions the magnetic field and plasma are ``frozen'' such that they move together\citep{Alfven42}. That is to say that ions and electrons may gyrate about fieldlines but they cannot move from one fieldline to the next. This is the typical situation within space plasmas where, on global scales, $R_m > 10^{8}$.

There are, however, highly localised regions where $0 < R_m \le 1$ known as diffusion regions. When the magnetic Reynolds number is less than one ($R_m<1$), the diffusion term dominates, and this may permit the magnetic field to slip through the plasma.
Reconnection can occur when the magnetic Reynolds number is of order one, i.e., where the advection and diffusion terms are of the same order.

Although, magnetic reconnection is a local process it has important global consequences because it is the fundamental plasma physics process by which the magnetic field may reconfigure. The re-configuring of the magnetic field through reconnection only occurs if it results in a decrease in the magnetic energy of the system with the lost magnetic energy converted into thermal energy, bulk plasma motions and also the acceleration of particles\citep{MagRecon00}. The division of the magnetic energy into these three other energies will be dependent on the properties of the plasma, but the details are not not well understood and are beyond the scope of this paper.

\subsection{Consequences of Ohm's Law}\label{sec:consequences Ohms}

In order to better understand why magnetic reconnection in three dimensions differs so much to two dimensional reconnection we first consider the nature of the elements of Ohm's law.

In two spatial dimensions, both the magnetic field and velocity are dependent on two spatial variables and have just two components, e.g.,
$\boldsymbol{B}(x,y,t)$ and $\boldsymbol{v}(x,y,t)$.
The electric field $\boldsymbol{E}$ is also only a function of $x$, $y$ and $t$, thus, from Faraday's law 
(Eqn~\ref{eqn:faradays}), it can only have one component, namely, $\boldsymbol{E} = E_z(x,y,t)\;\hat{\boldsymbol{z}}\,.$
Thus, in 2D, Eqn~\ref{eqn:gen-ohms} shows that $\boldsymbol{E}\cdot\boldsymbol{B} = \boldsymbol{N}\cdot\boldsymbol{B} \equiv 0.$ 

On the other hand, in three dimensions, $\boldsymbol{B}$, $\boldsymbol{v}$  
and $\boldsymbol{E}$ all have three components that can depend on all three spatial variables, hence, in general $\boldsymbol{E}\cdot\boldsymbol{B} = \boldsymbol{N}\cdot\boldsymbol{B} \neq 0.$

As explained in a pair of papers by Schindler and Hesse\citep{Schindler88,Hesse88}, a region of non-zero parallel electric field, where 
$$E_\| = \boldsymbol{E}\cdot\boldsymbol{B}/\|\boldsymbol{B}\|^2\,,$$ 
turns out to be both necessary and sufficient for reconnection in 3D. In particular, the rate of reconnection associated with a domain $\mathcal{D}$ within which there is an isolated diffusion region (region where $\boldsymbol{N}\neq \boldsymbol{0}$) is
$$\max_{\forall l\in\mathcal{D}}\left\{ \int_l \boldsymbol{E}_\|\;{\rm d}l \right\} \neq 0\;,$$
where $l$ is a fieldline and ${\rm d}l$ is a small element of the fieldline.

Given the requirement for non-zero $E_\|$ an obvious question arises as to whether, in 3D, reconnection can occur exactly at a null point where $\boldsymbol{B}=\boldsymbol{0}$. %\textbf{Although $E_\|$ is undefined exactly at $\boldsymbol{B}=\boldsymbol{0}$ it is defined everywhere in the limit $\boldsymbol{B}\rightarrow \boldsymbol{0}$.}
Pontin \& Priest\cite{PontinPriest2009} present 3 scenarios for 3D null point reconnection: spine-fan, torsional fan and torsional spine reconnection: in each scenario, the null point lies within the diffusion region with reconnection occurring all around it and possibly as it. 

Reconnection at magnetic separators is quite distinct. As determined by Stevenson \& Parnell\cite{Stevenson2015a}, the region of non-zero parallel electric field occurs away from the null points at the ends of the separators. Indeed, if the separator is long in length (e.g., on the sun separators can be many solar radii in length\cite{Platten2014}) then reconnection can occur a significant fraction of a solar radii from the null points at either end. As a result, separator reconnection occurs in a manner that is consistent with non-null 3D reconnection, or reconnection in a region with a ``guide field'' where the guide field is parallel to the separator.  The importance of the separator is that, as a boundary between four topologically distinct flux domains, transfer between domains as a result of separator reconnection can result in a global restructuring of the magnetic field. Furthermore, numerical experiments show that, the maximum rate of reconnection (determined by integrating along all fieldlines within the domain), is located along a separator.

\section{Numerical Experiments involving Magnetic Reconnection at Separators\label{sec:num-experiments}}

The first numerical resistive MHD experiment to reveal magnetic reconnection sites threaded by magnetic separators was presented in [\onlinecite{Haynes2007}] with parallel electric fields used to locate the sites of reconnection. Their numerical experiment initially had no separators but there were two null points, both of which lay on the domain's base. As the system was driven the two separatrix surfaces from the null points pushed up against each other causing a current layer to form which lay midway between the null points. At the same instant that the reconnection started, two separators and a new flux domain (bounded by the separatrix surfaces and the two new separators) were created. This process was called a separator bifurcation by [\onlinecite{Haynes2007}] but it would be more accurate to describe it as a separator-separator bifurcation as two separators are created (or destroyed) following the bifurcation. 

Two interesting findings from their work is that the integral of the parallel electric field along one of the two newly created separators was always of opposite sign to that along the other separator and the rates of reconnection at two new separators was different immediately after the bifurcation. The first finding indicated that the reconnection on the separators was in opposite directions. That is along one separator the reconnection acted to fill the newly formed flux domain that lay between the two separators and along the other it acted to empty this new domain. Since, the flux in the newly formed domain increased over time (at least initially) the reconnection which led to the filling of the domain ran at a faster rate than that which emptied the domain.

Another couple of papers relevant for understanding the creation of separators and their importance for reconnection\citep{Maclean2009,Parnell2010ApJ} studied a numerical resistive MHD experiment in which the magnetic reconnection, identified via parallel electric fields, was driven by the emergence of a twisted magnetic flux tube through the solar photosphere into a horizontal overlying magnetic field that filled the solar atmosphere. The rise of the flux tube was driven by convective motions and then a magnetic buoyancy instability kicked in enabling flux from the tube to rise up into the solar atmosphere\cite{Archontis2005}. The domain was initially devoid of magnetic null points in the solar atmosphere and, thus, had no topological features there either. Topological features were created as the experiment continued making it an ideal experiment in which to investigate the creation, behaviour and destruction of topological features.

\begin{figure}
  \centering
  \includegraphics[scale=1.1]{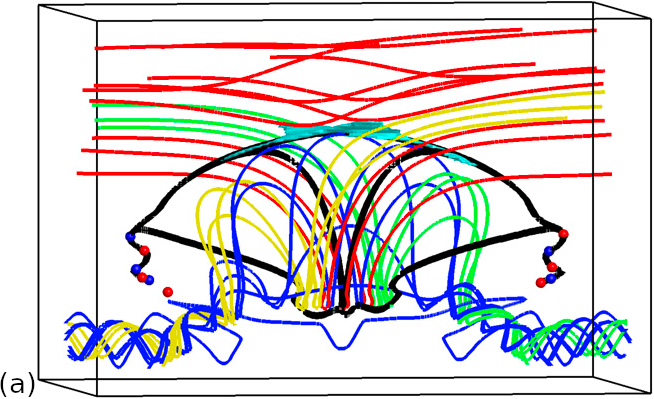}
  \includegraphics[scale=0.3]{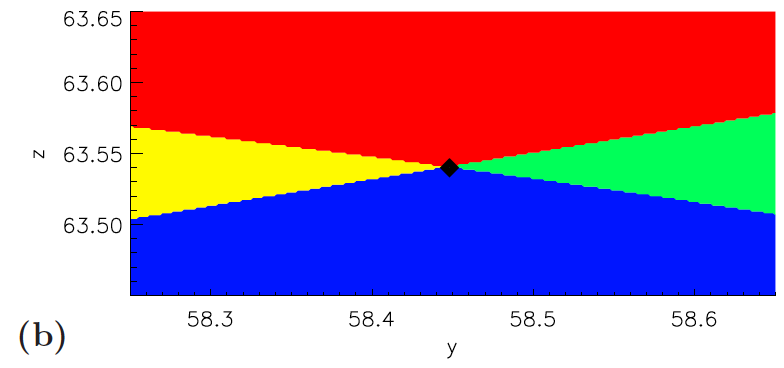}
  \includegraphics[scale=0.3]{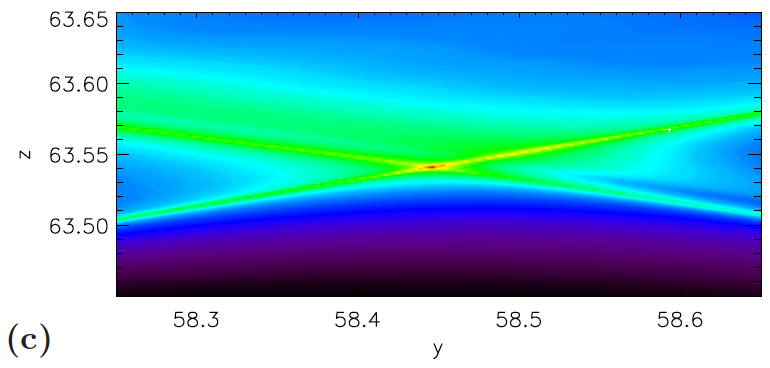}
  \caption{(a) A plot of the topological structure and fieldlines of one frame from the flux emergence experiment\citep{Parnell2010ApJ} showing the positive and negative null points (red and blue spheres) and the separators (thick black lines). (b) and (c) show a vertical cut in the $x=0$ plane of (a) located about the separator (black line) going through the current layer (cyan surface) located between the flux tube (blue lines) and overlying coronal field (red lines). (b) shows the connectivity of all the fieldlines that thread the cut with the diamond highlighting the separator which intersects the plane at the boundary between the 4 different types of magnetic field connectivity. In (c) the integrated parallel electric field along the fieldlines used to make (b) is plotted with low to high integrated parallel electric field indicated by deep purple/blue to red. Fig 1 from [\onlinecite{Parnell2010ApJ}].
  Reproduced with permission from Parnell, Maclean and Haynes, Astrophys. J. Lett., 725, L214 (2010). Copyright 2008, American Astronomical Society.} 
\label{fig:flux-emerge}
\end{figure}
Maclean et al.\cite{Maclean2009} identified and analysed the behaviour of the magnetic null points in this experiment. As the magnetic field from the flux tube emerged into the solar atmosphere, an initial region of weak magnetic field formed on one side of the outer edge of where the flux tube had arisen. Within this weak field region a pair of magnetic null points was formed via a null-null bifurcation.   
Approximately 6 minutes later, a second weak magnetic field region, containing a pair of newly created magnetic null points, was formed between the flux tube and the overlying field but this time lying on the opposite side to the other weak field region\citep{Maclean2009}. The two weak-field regions where the new null point pairs formed were called null clusters. 

During the experiment, [\onlinecite{Maclean2009}] showed that within each null cluster, pairs of opposite-polarity null points were created, and different pairs destroyed, on a regular basis with most nulls lasting only 8 frames or less (less than 2.5 minutes). There were, however, a few null points that lasted for longer with two null points lasting for the duration of the reconnection phase in the experiment. These two null points resided in different null clusters with one a positive null and the other negative. Each cluster always had an even number of null points in it: half being positive nulls and the other half negative nulls. 

One snapshot from this experiment is illustrated in Figure~\ref{fig:flux-emerge}a. The null points within this snapshot are denoted by the red and blue spheres that form two groups, one either side of the emerged flux tube. Each newly formed pair of null points consists of one positive (red sphere) and one negative (blue sphere) null. Moreover, [\onlinecite{Maclean2009}] found that as the flux tube rose higher into the coronal overlying field a current layer was formed between these two magnetic regions (cyan surface in Figure~\ref{fig:flux-emerge}a). The current layer (at times split into several pieces) lay at a higher height and midway between the two null clusters. At no time did the current layer encroach into the null clusters.

\begin{figure}
  \centering
  \includegraphics[scale=0.6]{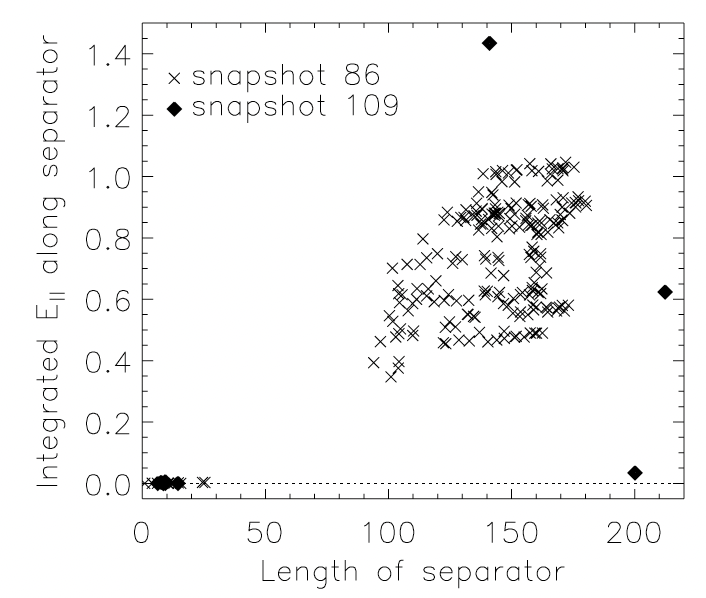}
  \caption{Plot of integrated parallel electric field along a separator versus length of the separator for separators found in two snapshots from the flux emergence experiment (snapshot 109, red diamonds, is the snapshot illustrated in Figure~\ref{fig:flux-emerge}). A unit of integrated parallel electric field equates to 150 V whilst a unit of length corresponds to 170 km. Fig 4 from [\onlinecite{Parnell2010ApJ}].
  Reproduced with permission from Parnell, Maclean and Haynes, Astrophys. J. Lett., 725, L214 (2010). Copyright 2008, American Astronomical Society.} 
\label{fig:flux-emerge-Epar}
\end{figure}

Considering the magnetic field in the same flux emergence experiment as that described above, [\onlinecite{Parnell2010ApJ}] determined the time evolution of its full magnetic topology (separatrix surfaces and spines of every null point and all the separators within the system). A number of surprising results were found. 
The null points within an individual null cluster are typically connected together by separators (short black lines in Figure~\ref{fig:flux-emerge}a), called cluster separators, such that they form a chain of null points and separators (like beads (null points) on a string (the connected separators)). 
Not all null points in a cluster are part of the chain, but most were. 
The longest lasting null in each cluster was typically located at one end of each chain. 
Finally, the integral of the parallel electric field was determined along each of the cluster separators and in their vicinity. All of these separators had an almost negligible integrated parallel electric field, indicating that little to no reconnection was associated with these separators, or null points (diamonds plotted in the bottom left-hand corner of Figure~\ref{fig:flux-emerge-Epar}).

 At the onset of reconnection between the coronal field and the flux tube (red and blue lines in Figure~\ref{fig:flux-emerge}a \& \ref{fig:flux-emerge}b, respectively), two new types of magnetic field connectivity were created: (i) magnetic field that entered the domain in the corona then became wound up in the flux tube and left the domain in the convective zone (green lines in Figure~\ref{fig:flux-emerge}a \& \ref{fig:flux-emerge}b) and (ii) magnetic field that entered through the convection zone as part of the flux tube and left as part of the coronal overlying field (yellow lines in Figure~\ref{fig:flux-emerge}a \& \ref{fig:flux-emerge}b). 
 
 The onset of reconnection possibly coincided with the formation of the first null pair, (null cluster) but so little reconnection occurred the results were inconclusive. On the creation of the second null pair (new null cluster), the reconnection was properly identified through the creation of green and yellow field lines. Also formed along with the second null cluster was a single separator connecting the null points that became the two long-lasting nulls which lay far apart in the separate null clusters. Shortly afterwards a multitude of separators were created connecting the same two long lasting null points\cite{Parnell2010ApJ}. The separators that connected null points from distinct null clusters are called ``intercluster separators'' (e.g., long black lines in Figure~\ref{fig:flux-emerge}a). After the first intercluster separator was formed, all the additional intercluster separators were formed in pairs. Similarly, different pairs of intercluster separators were also destroyed in pairs.

 Parnell et al.\citep{Parnell2010ApJ} found that during the most intense period of reconnection, many tens of intercluster separators (e.g., more than 200 in one snapshot) threaded the large intense current layer between the two null clusters and connected the same few null points. Practically all of these separators had a large integrated parallel electric field associated with them indicating they are sites of reconnection. The intercluster separators are much longer in length than those that linked nulls within each null cluster and some were fairly long lived. Moreover, when the parallel electric field was integrated along all fieldlines that threaded the midplane of the system (which was also the midplane of the current layer), it was found that there were multiple localised peaks in the integrated parallel electric field and each of these peaks coincided with an intercluster separator (Figure~\ref{fig:flux-emerge}c).
 
These numerical MHD studies, especially Parnell et al.\citep{Parnell2010ApJ}, revealed that not all separators seem to be sites of strong magnetic reconnection. Specifically, those that connect null points inside a null cluster (e.g. cluster separators) have an integral of parallel electric field along them that is small. However, intercluster separators that connect null points from two distinct null clusters can have, and often do have, a significant integral of parallel electric field (the largest such values in the domain) and they lie at the boundary between 4 globally-significant topologically-distinct flux domains indicating that they are important (the major) sites of reconnection (Figures~\ref{fig:flux-emerge}b \& ~\ref{fig:flux-emerge}c). 

In this paper, we consider whether the distinction between cluster and intercluster separators is just semantics or whether these features are intrinsically different. For the separators seen in Figure~\ref{fig:flux-emerge}a the cluster separators have lengths of $< 500 km$ and integrated parallel electric fields of $< 0.15$ V. The 3 intercluster separators are much longer with lengths of approximately $24$, $34$ and $36$ Mm and corresponding integrated parallel electric fields of $216.0$, $4.5$ and $93.8$ V, respectively (see red diamonds in Figure~\ref{fig:flux-emerge-Epar}). From an experiment run by Stevenson and Parnell\onlinecite{Stevenson2015a} the diffusion region extended along approximately half of the separator away from the nulls at the ends of the separator. Assuming this is the case for all of the separators mentioned in Figure~\ref{fig:flux-emerge}a, then mean parallel electric field values of $<6\times 10^{-4}$ V/km exist on the cluster separators and values of $\approx 1.8\times 10^{-2}$, $\approx 3\times 10^{-4}$ and $\approx 4\times 10^{-3}$ V/km on the 3 intercluster separators. Note, however, in reality, the strength of the parallel electric field may vary considerably along the length of a separator and the diffusion regions could be either short or longer than half the separator length. So, on two of the intercluster separators the mean parallel electric fields are factors of 30 and 6.7 times greater than the maximum values found on the cluster separators. Thus, the differences in their integrals of parallel electric field are not simply due to the differences in their lengths. 

\begin{figure}
  \centering
 \includegraphics[scale=0.8]{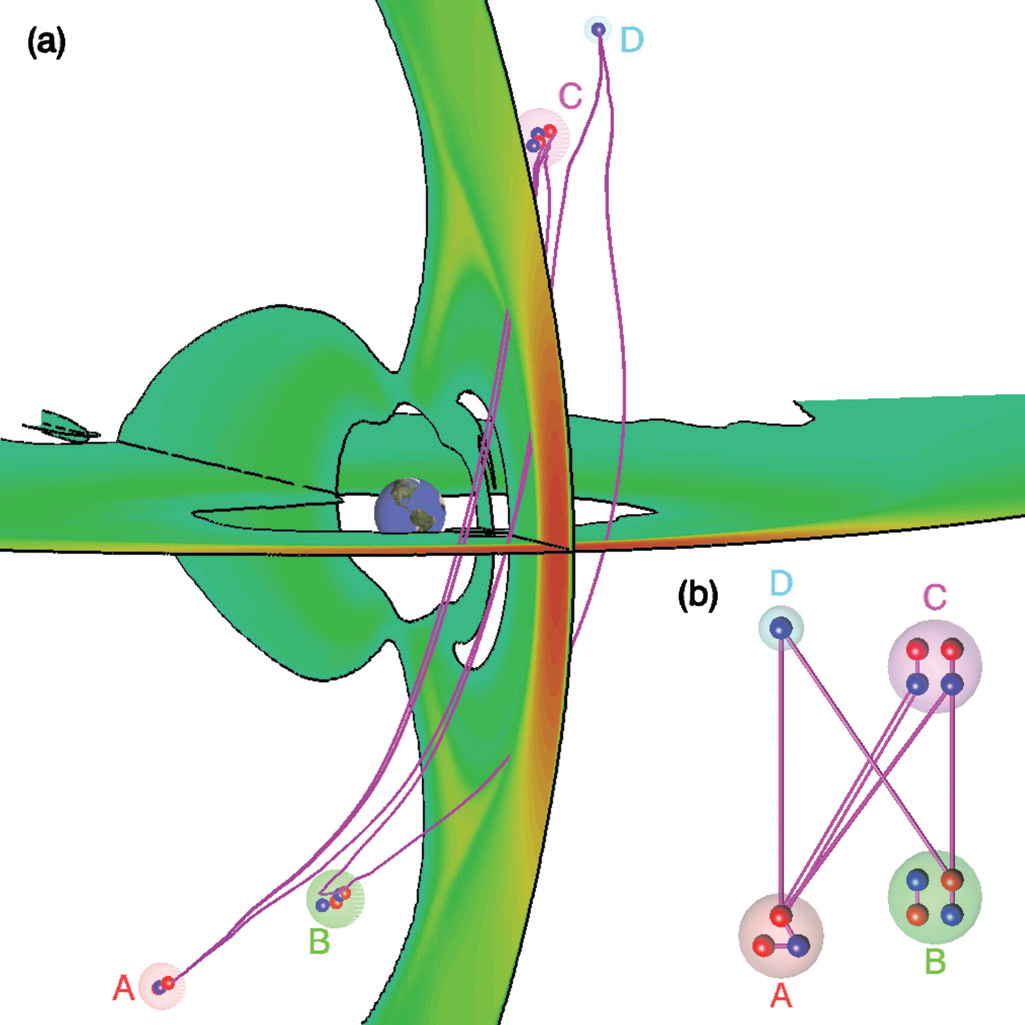}
  \caption{(a) The magnetic skeleton of the dayside magnetosphere with a northward interplanetary magnetic field at an angle of $\pi/4$. The null clusters, labelled A-D, contain positive (red) and/or negative (blue) null points. (b) A null-point connection map indicating the cluster separators (purple lines inside the shaded spheres) and the intercluster separators (long purple lines linking nulls from different clusters) that lie in the dayside magnetosphere. Fig 4 from [\onlinecite{Haynes2010}] Reproduced from Haynes and Parnell, Phys. Plasmas 17, 092903 (2010), with the permission of AIP Publishing.}
\label{fig:null-cluster-example}
\end{figure}
Another numerical experiment worth mentioning is one in which the Earth's magnetosphere is modelled using the GCCM code\cite{Dorelli2007}. In this experiment, the Earth's magnetic field interacts with a northwards interplanetary magnetic field (IMF) angled at $\pi/4$ with respect to the Earth's magnetic axis.  Figure~\ref{fig:null-cluster-example} shows a frame from this magnetosphere experiment\citep{Haynes2010}.

The initial magnetic field in this experiment starts with a pair of null points lying on the dayside magnetopause. Specifically a positive null lies in region A and a negative one in region D in Figure~\ref{fig:null-cluster-example}. As the experiment proceeds, additional nulls are created in the weak-field regions about the original null points forming null clusters. The null clusters A and D have topological degrees\cite{Deimling1985,Greene1988} of -1 and +1, respectively, equal to the topological degree of the null point residing in the region at the start of the experiment. In addition, another two null clusters are formed (B and C): both of these regions have zero topological degree as neither of these clusters/weak-field regions existed initially and the new null points are created in opposite-polarity pairs. 

A paper by Albright\cite{Albright1999} explained that new null points are created due to perturbations in regions of weak magnetic field strength, such as around pre-existing nulls. It is therefore not surprising that additional null points are created about the existing nulls in the magnetopause\cite{Dorelli2007,Haynes2010}  as well as additional nulls forming in two new regions. The additional nulls are always created in opposite polarity pairs.

In Figure~\ref{fig:null-cluster-example}, a total of 6 intercluster separators and 6 cluster separators are found (purple lines). Four of the intercluster separators are single separators each connecting nulls lying in different null clusters. The other two separators connect the same two null points that lie in null clusters A and C. 

Null clusters may contain one or more nulls within them with the overall topological degree of a cluster determined by summing up the topological degree of all the nulls within the cluster. Although the number of null points within a cluster may change over time, the topological degree of each cluster is preserved unless (i) two null clusters coalesce to form a single cluster whose topological degree is then equal to the sum of the degrees of the two original clusters or (ii) a single null cluster divides into two separate clusters with the sum of the topological degree of the two new clusters equal to the topological degree of the original cluster.

From a distance, the magnetic field about a null cluster with a topological degree of +1(-1) will look like a negative(positive) null point. A region with topological degree of 0 will, from a distance, have a magnetic field that has no obvious appearance of containing null points, indeed, it may, look like a region of uniform magnetic field.

Separators can be created by one of two types of bifurcation: namely a null-null bifurcation\cite{Lau1990} (also known as a local separator bifurcation\cite{BrownPriest1999})  or a separator-separator bifurcation\cite{Haynes2007}. As the name of the former suggests, two null points are formed during a null-null bifurcation and a separator may also be formed connecting the newly formed nulls. The latter bifurcation leads to the creation of a pair of separators. 

In order to investigate cluster and intercluster separators kinematic models are considered in this paper in order to try and explain the key behaviours seen in the numerical experiments discussed above. In particular, what are the mechanisms by which these types of separator can form? Can simple time-dependent magnetic field models shed any light on why the short cluster separators are associated with weaker parallel electric field than the long intercluster separators and thus result in less reconnection?

\section{Null-null Bifurcations}\label{sec:creation of nulls}
\subsection{Local Field about 3D Nulls}
The local magnetic field about a null located at $\boldsymbol{x}_0 = (x_0,y_0,z_0)$ may be revealed by linearising the magnetic field about the null\citep{Parnell1996} leading to
 \begin{eqnarray}
 \boldsymbol{B} =  \boldsymbol{M}\cdot (\boldsymbol{x}-\boldsymbol{x}_0), 
 \label{eqn:local-null-field}
 \end{eqnarray}
 where $\boldsymbol{M}$ is a real non-singular $3\times 3$ matrix (the Jacobian matrix of the vector field $\boldsymbol{B}$ evaluated at $\boldsymbol{x}=\boldsymbol{x}_0$). This means the $Tr(\boldsymbol{M})=0$ since $\boldsymbol{B}$ must satisfy the solenoidal constraint. Moreover, this implies that the sum of the eigenvalues of $\boldsymbol{M}$ is also zero. If the real part of two of the eigenvalues is positive then the null point is said to be a positive (B-type) null. The eigenvectors associated with these eigenvalues define the separatrix surface local to the null point within which lie fieldlines that are all directed out from the null (pink lines in Figure~\ref{fig:posnull-sep}a). The remaining eigenvalue will be real and negative in sign. Its eigenvector defines the spines local to the null point: these are a pair of fieldlines that in this case are directed in towards the null point (blue lines in Figure~\ref{fig:posnull-sep}a). On the other hand, if the signs of the real parts of two eigenvalues are negative then the null is a negative  (A-type) null point with the associated eigenvectors defining the separatrix surfaces. In this case, the fieldlines in the surface are all directed in towards the null point. The eigenvector associated with the single positive eigenvalue defines the spines which are a pair of outward-directed fieldlines from the null.

As a brief aside, we note that the only 3D magnetic null points that can exist within a magnetohydrostatic equilibrium magnetic field are potential nulls. This was demonstrated in [\onlinecite{Parnell1997}] by taking the curl of the magnetohydrostatic momentum equation in the absence of gravity and viscosity which gives 
\begin{eqnarray*}
    \boldsymbol{\nabla}\times\left(-\boldsymbol{\nabla}p + \boldsymbol{j}\times\boldsymbol{B}\right) &=& \boldsymbol{\nabla}\times(\boldsymbol{j}\times \boldsymbol{B}) \\  &=&\boldsymbol{j}\cdot\boldsymbol{\nabla}\boldsymbol{B}
    = \boldsymbol{j}\cdot\boldsymbol{M} = \boldsymbol{0},
\end{eqnarray*}
because $\boldsymbol{j}$ is a constant since $\boldsymbol{B} = \boldsymbol{M}(\boldsymbol{x}-\boldsymbol{x}_0)$. 
However, since at a 3D null point the matrix $\boldsymbol{M}$ is non-singular, the above can only be satisfied if $\boldsymbol{j}=\boldsymbol{0}$. A magnetohydrodynamic (MHD) equilibrium can contain non-potential 3D null points as the Lorentz forces of such null points can be balanced by an appropriate flow\citep{Titov2000}. 

The kinematic bifurcation models here are not in equilibrium and so the null points in these models have non-zero currents at their null points.

\subsection{Null-Null Bifurcation Magnetic Field}\label{sec:null-null-cluster-separator-only}

A number of authors have considered null-null bifurcations (for example, [\onlinecite{Lau1990,BrownPriest1999,Murphy2015}]). Lau \& Finn\cite{Lau1990} claimed that a new null pair would always be connected by at least one separator. Murphy et al.\cite{Murphy2015} presented 3 examples of null-null bifurcations. Here, a kinematic model is presented in which a time-dependent magnetic field is prescribed from which both the electric field and plasma velocity perpendicular to the magnetic field can be determined. The magnetic field that is prescribed is chosen with the aim of determining all the possible local magnetic field configurations that might arise due to the emergence of a new pair of null points: this has not been done in the other papers. Since the magnetic field is local only terms up to and including order two are considered.

To determine the local magnetic field configurations the possible transition states are first identified. From Eqn~\ref{eqn:local-null-field}, it can be seen that 3D nulls are structurally stable provided that $\bf{M}$ remains non-singular\citep{Murphy2015}. As discussed by [\onlinecite{Murphy2015}] when $\bf{M}$ becomes singular the 3D null is replaced by a degenerate null which is structurally unstable and exists only instantaneously as a transition between topologically stable states: one state with no null points and the other with two 3D null points or vice versa. 

The most common degenerate nulls have rank$({\bf M})=2$. 
These nulls have two non-zero eigenvalues summing to zero and thus either (i) two real eigenvalues of the same magnitude but opposite in sign or (ii) a purely imaginary complex-conjugate pair. 
A third possibility is that the degenerate null has rank$({\bf M})=0$  so the matrix ${\bf M}$ has 3 zero eigenvalues. 

When a new pair of 3D null points is created through a null-null bifurcation at the origin a degenerate null must first form at that point. The nature of the degenerate null determines the types of 3D nulls created and thus the nature of the null-null bifurcation. A null-null bifurcation containing a degenerate state of rank$({\bf M})=0$ is less likely to occur than either of the other two cases which have rank$({\bf M})=2$ as the former requires all the eigenvalues of the linear field to be zero at the same time. It is included, though, for completeness. 

In order to illustrate the different types of null-null bifurcation we consider the time-dependent magnetic field of the form
\begin{eqnarray}
\label{eqn:null-null-bifur}
{\bf B}({\bf x}) + \delta {\bf B}({\bf x},t)  = \left[\begin{array}{c}
(a-z)x+by \\  cx-(a+z)y \\ z^2 \end{array}\right] + \left[\begin{array}{c}
\alpha x - \beta y \\ \gamma x - \alpha y  \\ - \delta^2 \end{array}\right]t,
\end{eqnarray}
where $a$, $b$, $c$, $\alpha$, $\beta$, $\gamma$ and $\delta$ are constants and $t$ can be thought of as a relative time with $t=0$ corresponding to the instant of the null-null bifurcation. This magnetic field is similar to that considered in [\onlinecite{Murphy2015}] who only investigated null-null bifurcations involving degenerate nulls of rank$({\bf M})=2$. 

The form of the magnetic field is chosen as it is the simplest form that can produce all the possible degenerate null states and all the subsequent types of 3D null point pairs that can arise from them (excluding topologically equivalent states arising from rotations or reflections of the individual nulls).

It is noted here that the electric field $\boldsymbol{E}$ can be calculated for this magnetic field using Eqn~\ref{eqn:faradays}. From these fields the plasma velocity perpendicular to the magnetic field can then be determined assuming ideal evolution (i.e., assuming $\boldsymbol{N}=\boldsymbol{0}$). The electric field and perpendicular velocity are calculated in Section~\ref{sec:separator reconnection}. In this section, the focus is on the magnetic field so each different null-null bifurcation can be determined.

Turning to the form of the magnetic field, Eqn~\ref{eqn:null-null-bifur}, it may be noted that the $B_z$-component can only vanish if $t\ge 0$. Thus, when $t<0$ there are no null points in the magnetic field.

\subsubsection{Moment of Bifurcation}
At $t=0$ an unstable degenerate null, located at $(0,0,0)$, is momentarily created with eigenvalues 
$$\lambda_{1,2} = \pm\sqrt{a^2+bc}, \;\; \lambda_3 = 0\;.$$
The associated eigenvectors are
\begin{equation}
        {\bf e}_1 \equiv 
            \begin{bmatrix}
                -b \\ a-\sqrt{a^2+bc} \\ 0
            \end{bmatrix}, \;\;
        {\bf e}_2 \equiv 
            \begin{bmatrix}
                -b \\ a+\sqrt{a^2+bc} \\ 0
            \end{bmatrix}  \; \mbox{and } \;
        {\bf e}_3 \equiv 
            \begin{bmatrix}
                0 \\ 0 \\ 1
            \end{bmatrix}. \nonumber
\end{equation}
Depending on the relative values of the constants $a$, $b$ and $c$ the degenerate null may have one zero eigenvalue and (i) two real eigenvalues ($a^2+bc>0$), (ii) a complex conjugate pair of eigenvalues ($a^2+bc<0$) or (iii) an additional two zero eigenvalues ($a^2+bc=0$).

\subsubsection{New Pair of 3D Nulls}
\begin{table*}
\caption{\label{tab:new-nulls} Null-null bifurcation examples.}
\begin{ruledtabular}
    \begin{tabular}{llccc}
        Examples & &  1) $\;\;a^2+bc>0$ &  2) $\;\;a^2+bc=0$ &  3) $\;\;a^2+bc<0$  \\
        \hline
%        & $a,b,c$ & $2,-1,3$ & $1,1,-1$ & $1,1,-2$ \\
        & & & &\\
 %       Times & & & &\\       
        $t<0$ & & No nulls & No nulls & No nulls \\
        & & & & \\
       $t=0$ & E-values & real, equal, opposite sign & all equal to zero & complex conjugate pair \\
       & Magnetic field & hyperbolic (X-type) & anti-parallel lines & elliptic (O-type) \\
       & & & &\\
       $0<t\ll 1$ & E-values\footnote{Assume $0 < \delta \sqrt{t} < 1$ and $|h(t)|\ll |a^2+bc|$} & $\lambda_{1,2} = - \delta \sqrt{t} \pm\sqrt{a^2+bc + h(t)}$  & $\lambda_{1,2} = -\delta \sqrt{t} \pm \sqrt{h(t)}$ & $\lambda_{1,2} = -\delta \sqrt{t}
       \pm i\sqrt{h(t) -(a^2+bc)}$ \\
       & & $\lambda_3 = 2\delta \sqrt{t}$ & $\lambda_3 = 2\delta \sqrt{t}$ & $\lambda_3 = 2\delta \sqrt{t}$ \\
        &  & 3 real and distinct & 3 possibilities & 1 real and two complex conjugate \\
        & 3D Nulls &  & $h(t)< 0\;\;\Rightarrow\;\;$ Spiral &  \\
        &  & Improper & $h(t)= 0\;\;\Rightarrow\;\;$ Critical Spiral & Spiral \\
        &  &  & $h(t)> 0\;\;\Rightarrow\;\;$ Improper & \\
%       (a)  & $\delta\boldsymbol{B}(\boldsymbol{x},t)$ & $[t,-t,-t]^T$ & $[t,t,-2t]^T$ & $[t,t,-2t]^T$ \\        
%       & $\boldsymbol{x}_n(t)$\footnote{Coordinates of the new null point pair.} & $\left(\frac{t(3 \pm \sqrt{t})}{t-1},\frac{t(5 \mp \sqrt{t})}{t-1}, \pm \sqrt{t}\right)$  & & $\left(\frac{t(2 \pm \sqrt{2t})}{2t+1},\frac{-t(3 \mp \sqrt{2t})}{2t+1}, \pm \sqrt{2t}\right)$ \\
%       & ${\displaystyle{\lim_{t\rightarrow 0^+}\dot{\boldsymbol{x}}_n(t)}}$\footnote{Velocity vector of the new nulls at the instance of creation.} & $(-3,-5,\pm\infty)$ & & $(2,-3,\pm\infty)$ \\
%       & Spine & $\hat{\boldsymbol{e}}_1$ & $\hat{\boldsymbol{e}}_2$ & $\hat{\boldsymbol{z}}$ & $\hat{\boldsymbol{z}}$ & & \\
       & Separator &  & $h(t)<0$ - No  & \\
       &  & Yes & $0\le h(t)<\delta\sqrt{t}$ - No  & No \\
       &  &  & $0<\delta\sqrt{t}<h(t)$ - Yes  &  \\
%       & & & &\\
%       (b) & $\delta\boldsymbol{B}(\boldsymbol{x},t)$ & $[t,0,3ty-2t]^T$ & $[t,0,3yt-2t]^T$  & $[t,0,3yt-2t]^T$ \\
%        & $\boldsymbol{x}_n(t)$\footnote{Coordinates of the new null point pair.} & $\left(\frac{t(3 \pm \sqrt{t})}{t-1},\frac{t(5 \mp \sqrt{t})}{t-1}, \pm \sqrt{t}\right)$  & & $\left(\frac{t(2 \pm \sqrt{2t})}{2t+1},\frac{-t(3 \mp \sqrt{2t})}{2t+1}, \pm \sqrt{2t}\right)$ \\
%       & ${\displaystyle{\lim_{t\rightarrow 0^+}\dot{\boldsymbol{x}}_n(t)}}$\footnote{Velocity vector of the new nulls at the instance of creation.} & $(-3,-5,\pm\infty)$ &  &  $(1,-2,\pm\infty)$ \\
%       & Spine & $\hat{\boldsymbol{e}}_1$ & $\hat{\boldsymbol{e}}_2$ & $\hat{\boldsymbol{z}}$ & $\hat{\boldsymbol{z}}$ & & \\
%       & Separator & Yes & Yes & Yes
       \end{tabular}
\end{ruledtabular}
\end{table*}
When $t\ge 0$ a pair of generic 3D nulls are formed of opposite sign that are located at 
$${\boldsymbol{x}}_n(t) = (x_n(t), y_n(t), z_n(t)) = (0,0,\pm \delta \sqrt{t}).$$ The speed of separation of this new pair of null points is given by
$$\frac{{\mathrm{d}}\boldsymbol{x}_n}{{\mathrm{d}}t} = \left(0,0,\pm \frac{\delta}{2\sqrt{t}}\right),$$ 
which implies that, at the instant of creation, the new null points move apart at infinite speed (in the direction of (one of) the eigenvector(s) associated with the zero eigenvalue of the degenerate null)\citep{Murphy2015}. This means that, when first observed, a new pair of null points will likely be some distance apart. 

The nature of the newly created 3D null points is dependent on the sign of $a^2+bc$ as detailed in Table~\ref{tab:new-nulls}. 
This is because, after the bifurcation, the eigenvalues of the newly formed 3D null points will be perturbations of the eigenvalues $\pm\sqrt{a^2+bc}$ at the instant of bifurcation. Specifically, at the newly formed 3D null point located at $(0,0,z_n)$ the eigenvalues will be of the form
\begin{equation*}
    %\label{eqn:3Dnull-evalues}
    \lambda_{1,2} = -\delta\sqrt{t} \pm\sqrt{a^2+bc+h(t)}\;, \quad \mathrm{and} \quad \lambda_{3} = 2\delta\sqrt{t}\;,
\end{equation*}
where $h(t) = (2a\alpha +b\gamma -c\beta)t+(\alpha^2 - \beta\gamma)t^2\;.$
Since we are only interested in the time around the creation of the nulls we may assume that $t$ is sufficiently small such that $$|h(t)| < |a^2+bc| \quad {\mathrm{and}} \quad \delta^2 t < |a^2+bc|.$$
Moreover, the eigenvectors associated with these eigenvalues will also only be slight perturbations of the degenerate null's eigenvectors. 

In the sections below we consider in turn the three types of degenerate null and investigate the nature of all the possible 3D null point pairs that may arise following a null-null bifurcation via a transition state involving the given degenerate null. 

\subsubsection{Improper Nulls ($a^2+bc>0$)}
If $a^2+bc>0$ (Table~\ref{tab:new-nulls}, Example 1), the degenerate null formed at $t=0$ is an X-type 2D null point, lying in the $z=0$ plane. Once $t>0$ the two stable generic 3D nulls are formed. 

The 3D null at $(0,0,\delta\sqrt{t})$ has a positive eigenvalue $\lambda_3$ with eigenvector $\boldsymbol{e}_3$. Furthermore, since $h(t)$ is small we still find $a^2+bc+h(t)>0$. Moreover, $\delta^2 t < a^2+bc+h(t)$ and so the eigenvalues $\lambda_{1,2}$ will be of opposite sign. So this 3D null will be a positive null with a vertical separatrix surface defined by the eigenvectors $\boldsymbol{e}_1$ and $\boldsymbol{e}_3$ and horizontal spines parallel to $\boldsymbol{e}_2$. Similarly, we can argue that the 3D null at $(0,0,-\delta\sqrt{t})$ will be a negative null with a vertical separatrix surface defined by the eigenvectors $\boldsymbol{e}_2$ and $\boldsymbol{e}_3$ and horizontal spines parallel to $\boldsymbol{e}_1$.

\begin{figure}
  \centering
 \includegraphics[scale=0.31]{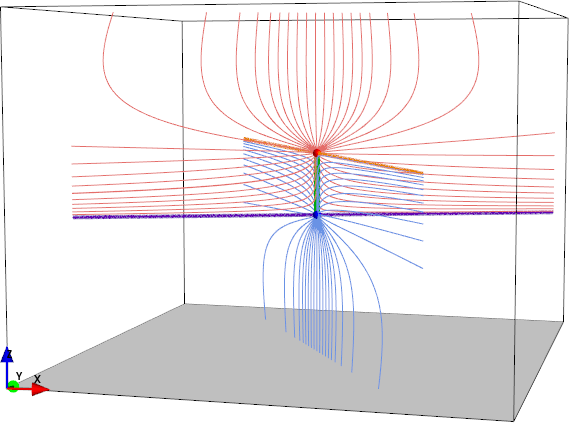}
 \includegraphics[scale=0.33]{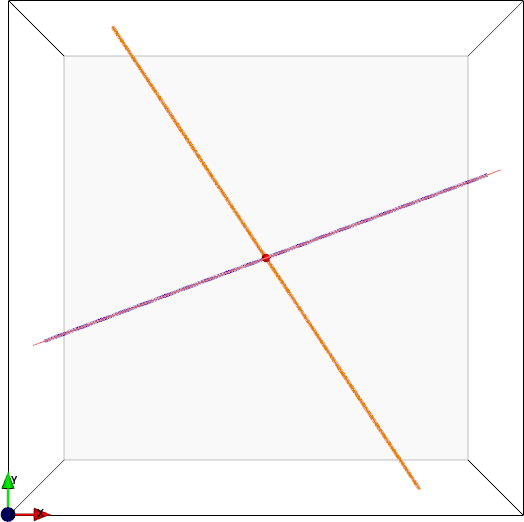}
  \caption{Plots showing the magnetic skeleton (viewed from the side and from above) following a null-null bifurcation starting from a degenerate null with real eigenvalues and forming two improper nulls (red and blue spheres) connected by a separator (green line). The separatrix surfaces of the positive and negative nulls are made up of red and blue lines with the spines of the nulls orange and purple, respectively. Here, $(a,b,c)=(1,2,1)$, $(\alpha, \beta, \gamma) = (1,1,1)$, $\delta=1$ and $t=0.3$.}
\label{fig:new-nulls-ex1}
\end{figure}
Since all the eigenvalues for both new 3D nulls are real they will both be improper nulls\citep{Parnell1996} (Figure~\ref{fig:new-nulls-ex1}). The topological degree of the region remains unchanged from before the bifurcation where it is zero, to after the bifurcation where the sum of the topological degree of the positive and negative null is also zero. 

Since the separatrix surfaces of the two nulls are both (non-parallel) vertical planes and the $z$-axis lies in both, it is clear that these surfaces must intersect. Indeed it can be seen that the positive null's separatrix surface is bounded by the spine of the negative null (red lines and thick purple line, respectively in Figure~\ref{fig:new-nulls-ex1}), whereas the negative null's separatrix surface is bounded by the spine of the positive null (blue lines and thick orange line, respectively in Figure~\ref{fig:new-nulls-ex1}). Moreover, there exists a separator, which lies at the intersection of the two separatrix surfaces (thick green line, respectively in Figure~\ref{fig:new-nulls-ex1}). Indeed, the separator is a fieldline that extends from the positive null to the negative null and it lies at the boundary between four flux domains.

\subsubsection{Spiral Nulls ($a^2+bc<0$)}
\begin{figure}
  \centering
 \includegraphics[scale=0.31]{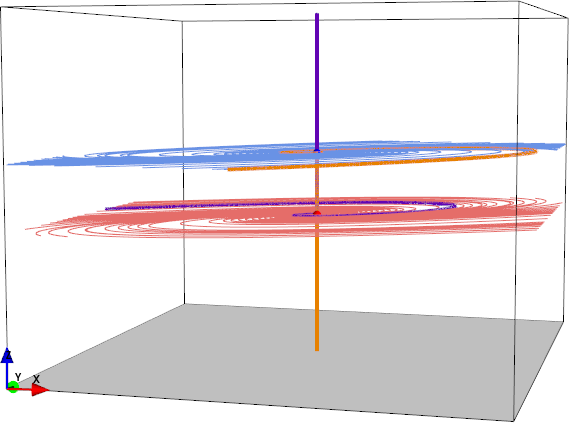}
 \includegraphics[scale=0.33]{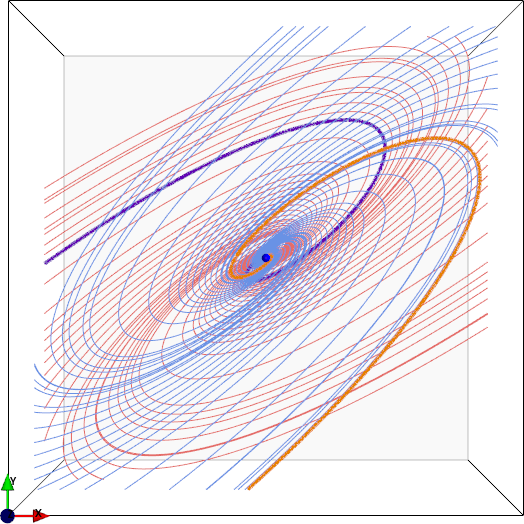}
  \caption{Plots showing the magnetic skeleton (viewed from the side and from above) following a null-null bifurcation starting from a degenerate null with purely imaginary eigenvalues and forming two spiral nulls that are not connected by a generic separator. Here, $(a,b,c)=(1,-2,1)$, $(\alpha, \beta, \gamma) = (1,0,1)$, $\delta=1$ and $t=0.3$. Topological features coloured as in Figure~\ref{fig:new-nulls-ex1}. }
\label{fig:new-nulls-ex3}
\end{figure}
If $a^2+bc < 0$ (Table~\ref{tab:new-nulls}, Example 3), then the degenerate null formed at $t=0$ is of O-type (it has two imaginary eigenvalues) that are lying in the $z=0$ plane as before. 

In Example 3, immediately following the null-null bifurcation two stable generic 3D nulls are formed both of spiral-type since $|h(t)| < |a^2+bc|$ (Figure~\ref{fig:new-nulls-ex3}). Thus, the eigenvalues $\lambda_{1,2}$ of the 3D nulls are a complex conjugate pair. For both null points, these eigenvalues are associated with eigenvectors that lie in a $z=$ constant plane. This means that locally the separatrix surfaces of both nulls are approximately parallel to each other. So, unless these surfaces curve in towards each other further from the nulls then no separator connects these null points.

The pairs of spines from each of these nulls are directed vertically upwards and downwards out of the nulls. The upwardly directed spine from the negative null and the downwards directed spine from the positive null coincide and thus they form a non-generic separator that will disappear with a small disturbance (such as the propogation of a wave in the domain). A more likely (generic) null-null bifurcation forming two spiral nulls would still be such that the separatrix surfaces of the new nulls were (at least initially) parallel but their spine lines would not coincide, as seen in Fig 4 in [\onlinecite{Murphy2015}].

The overall topological degree of the region remains unchanged, maintaining a value of zero throughout the null-null bifurcation.

In [\onlinecite{Murphy2015}] a case with two spiral nulls that were connected by a separator was discussed. It should be noted that in such a case it is always possible to choose a sufficiently short time $t$ after the bifurcation for which the two nulls are not connected by a separator within a given sized box about the nulls. This means a null-null bifurcation in which a pair of 3D spiral nulls are created may be connected by a generic separator if, further from the new null-point pair, the separatrix surfaces of the new nulls are angled in towards each other so they overlap.

\subsubsection{Critical Spirals ($a^2+bc=0$)}
In Table~\ref{tab:new-nulls} Example 2, $a^2+bc=0$ and at $t=0$ the degenerate null is of rank($\boldsymbol{M}$)=0. It is therefore not a 2D null point. 
This configuration, which only exists for an instant, permits a much greater degree of freedom (as it has three zero eigenvalues) and so the bifurcation can lead to various scenarios. 

In this case the two newly formed stable generic 3D nulls will have eigenvalues of the form
\begin{equation}
    \label{3Dnull-evalues2}
    \lambda_{1,2} = \mp\delta\sqrt{t} \pm\sqrt{h(t)}\;, \quad \mathrm{and} \quad \lambda_{3} = \pm 2\delta\sqrt{t}\;.
\end{equation}

\begin{figure}
  \centering
 \includegraphics[scale=0.31]{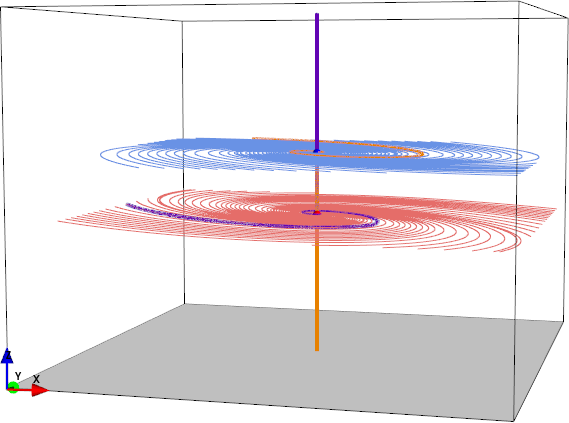}
 \includegraphics[scale=0.3]{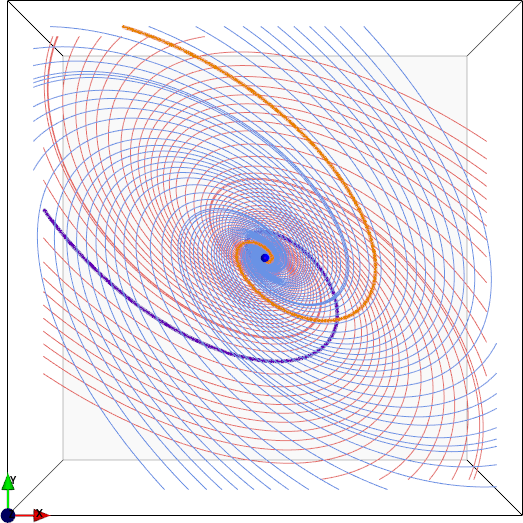}
  \caption{Plots showing the magnetic skeleton (viewed from the side and above) following a null-null bifurcation starting from a degenerate null with 3 zero eigenvalues. Example 2a (no separator): $t=0.3$, $(a,b,c)=(1,1,-1)$, $(\alpha, \beta, \gamma) = (-1,2,-1)$ and $\delta=1$. Topological features coloured as in Figure~\ref{fig:new-nulls-ex1}.}
\label{fig:new-nulls-ex2a}
\end{figure}
\begin{figure}
  \centering
 \includegraphics[scale=0.31]{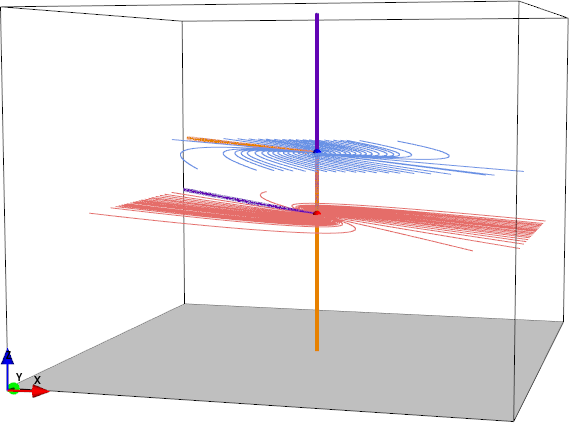}
 \includegraphics[scale=0.3]{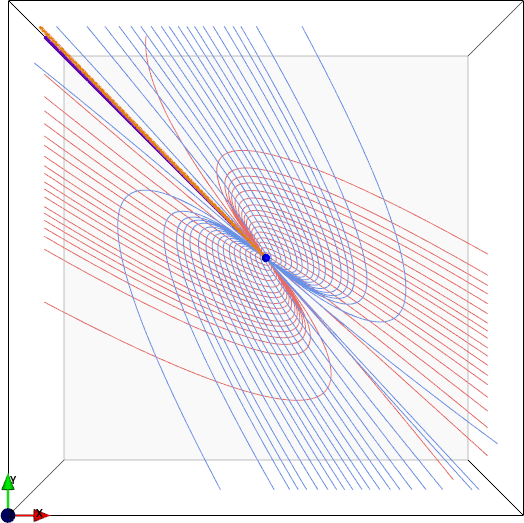}
  \caption{As for Figure~\ref{fig:new-nulls-ex2a} but here Example 2b (no separator): $t=0.3$, $(a,b,c)=(1,1,-1)$, $(\alpha, \beta, \gamma) = (-1,-1,1)$ and $\delta=1$.}
\label{fig:new-nulls-ex2b}
\end{figure}
\begin{figure}
  \centering
 \includegraphics[scale=0.31]{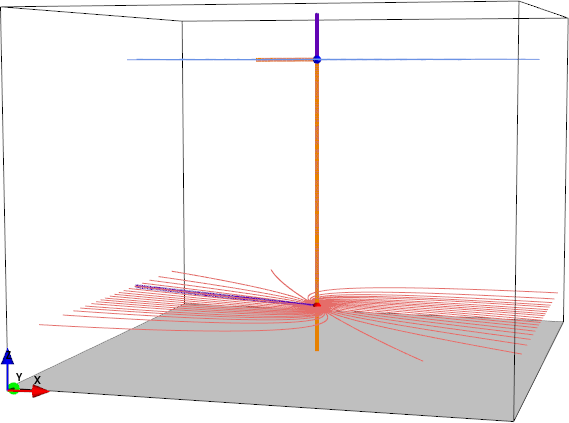}
 \includegraphics[scale=0.3]{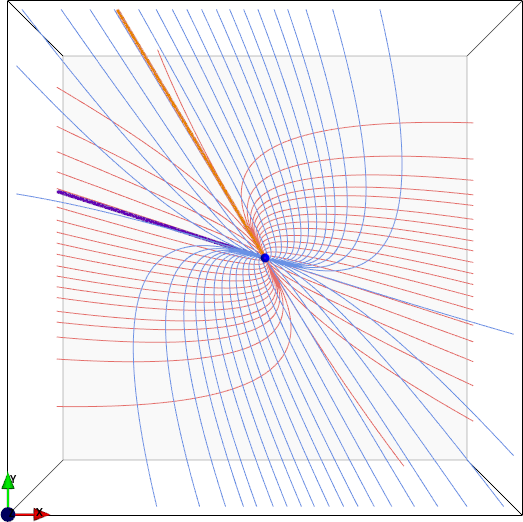}
  \caption{As for Figure~\ref{fig:new-nulls-ex2a} but here Example 2c (no separator): $t=0.3$, $(a,b,c)=(1,1,-1)$, $(\alpha, \beta, \gamma) = (1,1,1)$ and $\delta=4$.}
\label{fig:new-nulls-ex2cns}
\end{figure}
\begin{figure}
  \centering
 \includegraphics[scale=0.31]{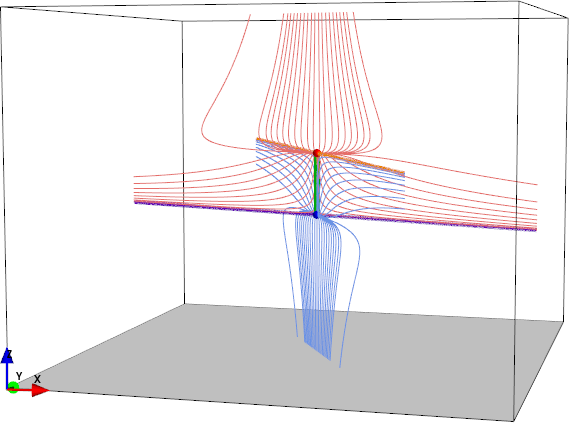}
 \includegraphics[scale=0.3]{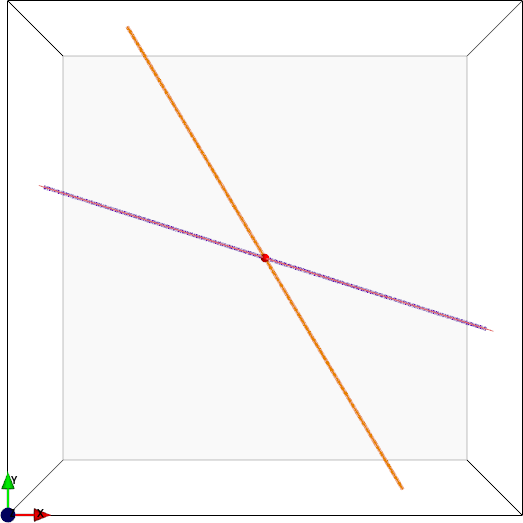}
 \caption{As for Figure~\ref{fig:new-nulls-ex2a} but here Example 2c (separator): $t=0.3$, $(a,b,c)=(1,1,-1)$, $(\alpha, \beta, \gamma) = (1,1,1)$ and $\delta=1$.}
\label{fig:new-nulls-ex2cs}
\end{figure}
When $h(t)<0$ two spiral nulls (one positive and the other negative) will be formed (Figure~\ref{fig:new-nulls-ex2a}). As in Example 2, the two nulls will not be connected by a separator unless the fan planes curve inwards towards the $z=0$ plane further from the null points themselves. 

In Figure~\ref{fig:new-nulls-ex2b}, the magnetic field is such that $h(t)=0$ so the nulls have two equal eigenvalues associated with horizontal separatrix surfaces of the nulls and thus the nulls are critical spirals which are not joined by a separator unless their separatrix surfaces curve in towards the $z=0$ plane further away from the nulls.

When $h(t)>0$ there will be two improper nulls. If $0\le h(t)<\delta^2 t$, then the two nulls will not be joined by a generic separator (Figure~\ref{fig:new-nulls-ex2cns}) unless the separatrix surfaces bend in towards each other outwith the linear field domain of each null but if $0\le \delta^2 t < h(t)$, the nulls will be connected by a separator (Figure~\ref{fig:new-nulls-ex2cs}).

As in the previous two cases, the overall topological degree of the domain remains zero at all times.

\subsection{Null-null bifurcation in a magnetic field with existing null points.\label{sec:null-null-intercluster-separator}}
All of the examples discussed above focus on the local consequences of a null-null bifurcation. Here, a kinematic model is presented in which a new pair of nulls are created in a magnetic field that already has existing null points. The distance between the emergence site of the new nulls and the location of the existing null pair does not influence the resulting configuration. So this model works just as well for a new null pair formed inside an existing null cluster as it does for a new null pair formed far away from the existing nulls in, or forming, a different null cluster.

The magnetic behaviour found in this model, which is just one example not an attempt to consider all possible such bifurcations, produces the same behaviour as that seen in the local-separator bifurcation described in [\onlinecite{BrownPriest1999}]. Here, though we still call this a null-null bifurcation as we view this as a possible additional consequence of such a bifurcation. 

The time-dependent magnetic field considered is
\begin{eqnarray}
\label{eqn:null-null-bifur-intercluster}
{\bf B}({\bf x})  = \frac{B_0}{L_0z_0z_1}\left[\begin{array}{c} 2x\left((z-a)^2(z+b) - bz_0^2t + az_1^2 +cz^2\right) \\
2y\left((z-a)(z+b)^2 - z(z_0^2t+z_1^2) -cz^2\right) \\
\left((z-a)^2 - z_0^2t\right)\left((z+b)^2 - z_1^2\right)
\end{array}\right]
\end{eqnarray}
where $a$, $b$, $c$, $z_0$, $z_1$, $B_0$ and $L_0$ are all positive constants and $t$ is a relative time with $t=0$ corresponding to the instant of the null-null bifurcation. 

As for the local null-null bifurcation model, both the electric field and plasma velocity component perpendicular to the magnetic field can be determined for this magnetic field. They are discussed in Section~\ref{sec:separator reconnection} where they are used to investigate reconnection about the separators created following the bifurcation.

\begin{figure}
 \includegraphics[scale=0.4]{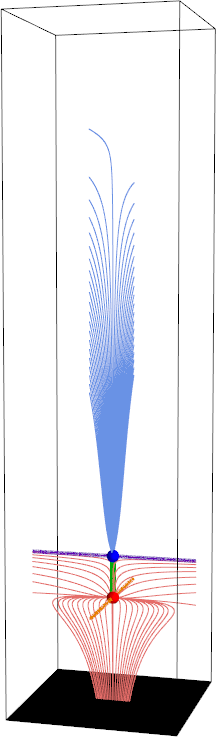}
 \includegraphics[scale=0.4]{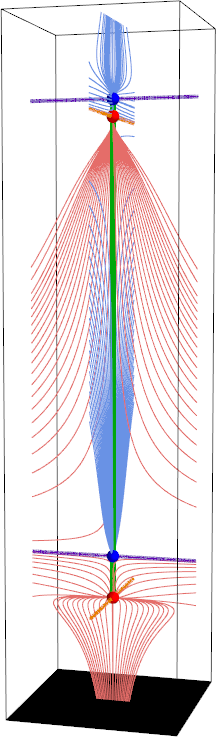}
  \includegraphics[scale=0.4]{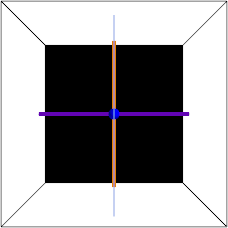}
 \includegraphics[scale=0.4]{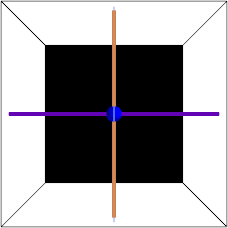}
  \caption{Plots showing the magnetic skeleton (viewed from the side (top) and above (bottom)) following a null-null bifurcation occurring within a magnetic field that already has some null points located at $-b\pm z_1 = -5\pm 0.5$, with $B_0=L_0=1$ and $c=3$. The left-hand snapshots are when $t=-1$ before the null-null bifurcation and the right-hand ones are after the bifurcation when $t=1$ and the new null pair are located at $a\pm z_0\sqrt{t}=6\pm 0.2$. Topological features are coloured as in Figure~\ref{fig:new-nulls-ex1}.}
\label{fig:new-nulls-inter-clust}
\end{figure}
Before the bifurcation ($t<0$) there exists a null pair located at $x=y=0$ and $z = -b\pm z_1$ with the lower null positive and the upper null negative (Figure~\ref{fig:new-nulls-inter-clust}(left)). These nulls are connected by a short separator. After the bifurcation ($t>0$) a new null pair have formed at $x=y=0$ and $z = a\pm z_0\sqrt{t}$: again the lower null is positive and the upper null negative (Figure~\ref{fig:new-nulls-inter-clust}(right)). Immediately following the bifurcation two new separators are formed: one connecting the new pair of nulls and the other connecting the new positive null point to the existing negative null. This new separator may be short, if the null pair is created close to the existing nulls, or long, if the new nulls are formed in a different null cluster to the existing nulls.

\section{Separator-Separator Bifurcation\label{sec:separators}}
As seen above, separators may be created along with a new pair of null points. 
%if the new nulls formed are improper (e.g. via a null-null bifurcation of type Example 1). Such a bifurcation creates a single separator that will initially be short in length as it connects the two newly formed null points that are created at a single point.
Separators may also be created in pairs via a separator-separator bifurcation\citep{Haynes2007,Parnell2007,Parnell2008,Parnell2010ApJ,Parnell2010JGR,Wilmot-Smith2011}. Note, when this bifurcation was first observed during the analysis of a numerical experiment that involved reconnection in the solar atmosphere it was called a separator bifurcation\citep{Haynes2007,Parnell2008}. Here, we call it a separator-separator bifurcation as it results in the creation of a pair of separators (or separator ring\citep{Longcope1996}), not just a single separator. It is essentially the same as the global-separator bifurcation described in [\onlinecite{BrownPriest1999}], however, in their magnetic field the second separator formed below the plane in which the point sources lay.  In such bifurcations the separatrix surfaces from two pre-existing, distant null points of opposite-sign approach one another. The bifurcation occurs when these separatrix surfaces intersect and a pair of separators are formed connecting the two existing null points. 

A kinematic model is presented below in order to illustrate how such a bifurcation can occur. Unlike the null-null bifurcation model (Section~\ref{sec:null-null-cluster-separator-only}), the model presented here only illustrates one separator-separator bifurcation rather than all possible such bifurcations. The time-dependent magnetic field considered is 
\begin{eqnarray}
\boldsymbol{B}({\bf x},t) &=& \frac{B_0}{L_0}\left[\begin{array}{c}
a(t)^2-x^2 \\  (2a(t)-1)xy/a(t) \\ (xz-a(t)b)/a(t)\end{array}\right] \nonumber \\
&& + \frac{c B_0}{L_0}\left[\begin{array}{c} 0 \\ 0 \\ (x^2+d^2)y^2 + (y^2+d^2)(x^2-a(t)^2)^2\end{array}\right]\;
\label{eqn:sep-bifur-field}
\end{eqnarray}
where $B_0>0$, $L_0>0$, $b$, $c$, $d$ are all constants and $a(t) \neq 0$ is a function of time.
The second part of the magnetic field has only a $z$-component and this term is positive provided $c>0$. It represents a vertical ambient field which becomes increasingly dominant the further $|x|$ is from $|a(t)|$ and the larger $|y|$ is.

As for the two null-null bifurcation models, the electric field and ideal plasma velocity perpendicular are calculated in the next section (Section~\ref{sec:separator reconnection}).

The magnetic field $\boldsymbol{B}(\boldsymbol{r},t)$ has two null points located at 
$$\boldsymbol{x}_{n1} = (a(t), 0, b) \;\; \mathrm{and} \;\; \boldsymbol{x}_{n2} = (-a(t), 0, -b).$$
The eigenvalues of the linearised field about these null points are
\begin{eqnarray*}
    \lambda_{n1} &\in & \left\{-2a(t),\; 2a(t)-1,\; 1\right\} \;\; \mathrm{and} \\
    \lambda_{n2} &\in & \left\{2a(t),\; 1-2a(t),\; -1\right\},
\end{eqnarray*}
and their associated eigenvectors are
$$(1,0,0)^T, \;\;\; (0,1,0)^T, \;\;\; (0,0,1)^T$$
respectively, in both cases.

For the separator-separator bifurcation to occur we require the following: (i) the null located at $\boldsymbol{x}_{n1}$ to be a negative null, (ii) the null at $\boldsymbol{x}_{n2}$ to be positive and (iii) the spines of both of these nulls to be parallel to the $z$-axis (i.e. to be aligned with the ambient field). This may be achieved by letting  $a(t)$ take a value between $0$ and $0.5$. Note, $a(t)\neq 0$ and if $a(t)=0.5$ then the null points are degenerate, so not generic 3D null points. We therefore define
$$a(t) = 0.5-t, \;\;\; 0<t<0.5\;.$$

\begin{figure*}
  \centering
  \includegraphics[scale=0.26]{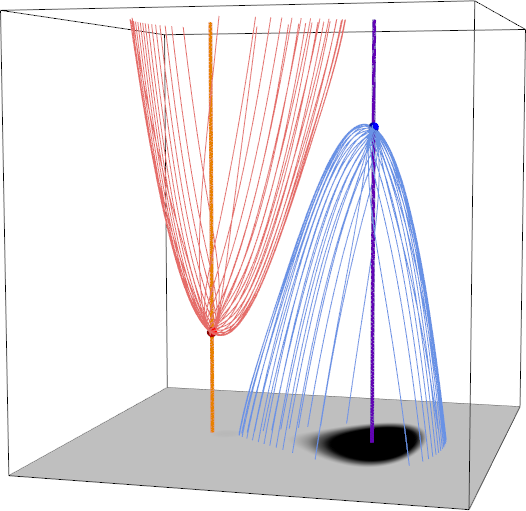}
  \includegraphics[scale=0.26]{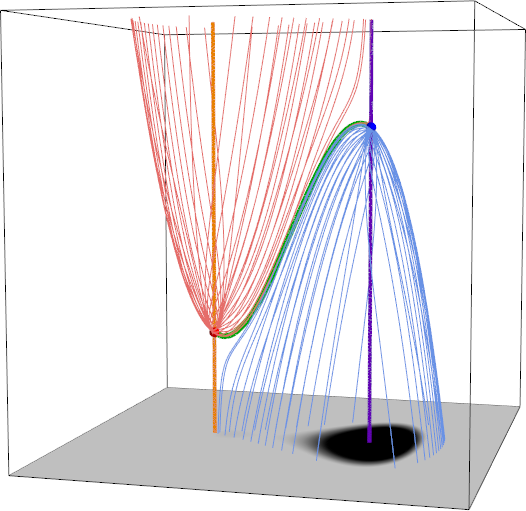}
  \includegraphics[scale=0.26]{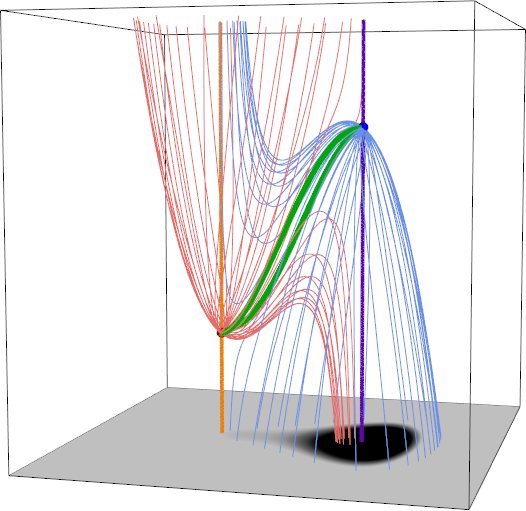}  
  \includegraphics[scale=0.26]{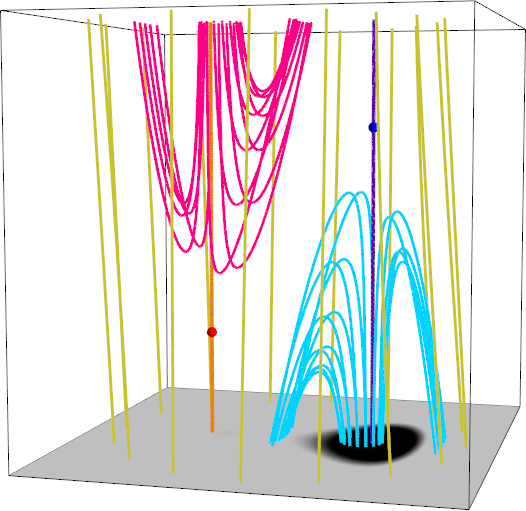}
  \includegraphics[scale=0.26]{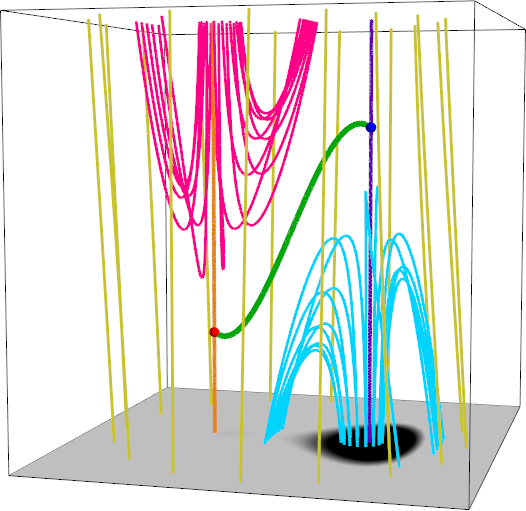}
  \includegraphics[scale=0.26]{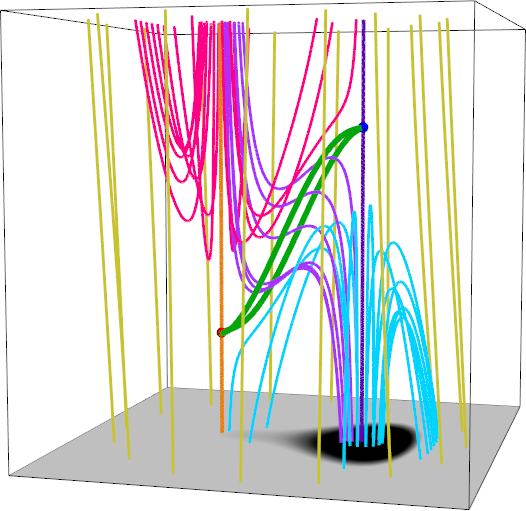} 
  \caption{Plots showing (top row) the skeleton of the magnetic field (left) before ($a(t)=0.4$), (middle) at the bifurcation ($a(t)=0.3874$) and (right) after ($a(t)=0.35$) a separator-separator bifurcation. The null points (large red/blue dots) and their associated spines (thick orange/purple lines) and fieldlines lying in their separatrix surfaces (thin red/blue lines) are plotted. The separators, when they exist, are plotted in dark green. The plots in the bottom row show general fieldlines at the same 3 times as those in the top row. Fieldlines (thick pink lines) lying above the positive separatrix surface, (thick cyan lines) lying below the negative separatrix surface, (thick yellow lines) the ambient vertical field and, in the final frame, (thick violet lines) lying above the positive separatrix surface and below the negative separatrix surface.}
\label{fig:sep-sep bifurcation}
\end{figure*}
The magnetic field plotted in Figure~\ref{fig:sep-sep bifurcation} has been constructed for the constants $B_0 = L_0 = 1$, $b=0.5$, $c=0.4$ and $d=5$.
If $0<t<0.113$ the separatrix surfaces of the two null points do not intersect and so there are no separators connecting the two null points (Figure~\ref{fig:sep-sep bifurcation}, left). Figure~\ref{fig:sep-sep bifurcation}(middle) shows the skeleton at the approximate time of the bifurcation ($t\approx 0.113$). At the instant of the bifurcation a single degenerate separator is located where the two separatrix surfaces first touch. As $t$ continues to increase ($0.113 < t< 0.5$) the two separatrix surfaces intersect each other forming two separators which move further apart in time. In the volume formed by the overlapping separatrix surfaces bounded by the separator ring a new flux domain is created which contains the thick violet coloured fieldlines seen in Figure~\ref{fig:sep-sep bifurcation}(right).

Having developed time-dependent magnetic field models for the bifurcations, the separators created in them are considered below to see if they are likely to be sites at which reconnection occurs.

\section{Separator Reconnection\label{sec:separator reconnection}}
 
Below we consider whether we can say anything about the reconnection associated with the separators formed in the three bifurcations. First, however, we recall that for 3D reconnection to occur
\begin{eqnarray*}
%\label{eqn:int-Epar}
    \int_l \frac{\boldsymbol{E}\cdot\boldsymbol{B}}{|\boldsymbol{B}|}\;{\rm d}l \neq 0\;,
\end{eqnarray*}
where, as before, $\boldsymbol{E}$ and $\boldsymbol{B}$ are the electric and  magnetic fields, respectively, and $l$ is fieldline length\citep{Schindler88,Hesse88}. 
%The rate of reconnection can be defined everywhere except when $\boldsymbol{B}\neq \boldsymbol{0}$.

Furthermore, the reconnection rate is defined as the maximum of all the integrals of the parallel electric field along the fieldlines. In the numerical experiments in which separators have been identified and the integrated parallel electric fields have been determined along fieldlines the local maxima all coincide with separators (e.g. [\onlinecite{Parnell2010ApJ}]). 

From our time-dependent bifurcation magnetic fields the electric fields may be calculated from Faraday's law (Eqn~\ref{eqn:faradays}) such that
$$\boldsymbol{E} = -\frac{\partial \boldsymbol{A}} {\partial t}  -\bnabla \phi \approx -\frac{\partial \boldsymbol{A}} {\partial t},$$ 
where $\boldsymbol{B} = \bnabla \times \boldsymbol{A}$, with $\boldsymbol{A}$ the magnetic vector potential and $\phi$ the electric potential. The above approximation is valid in a quasi-neutral plasma under the assumption that the length scales are much greater than the Debye length.
Thus, in each model the reconnection rate along each of the separators can be determined by using the above approximation for $\boldsymbol{E}$.

In addition, from Ohm's law (Eqn~\ref{eqn:gen-ohms}) the plasma velocity perpendicular to the magnetic field $\boldsymbol{v}_{\perp}$ can be determined
\begin{equation}
\label{eqn:vperp}
\boldsymbol{v}_{\perp} = \frac{(\boldsymbol{E}-\boldsymbol{N})\times\boldsymbol{B}}{|\boldsymbol{B}|^2}\,.
\end{equation}
Out with the diffusion region, the non-ideal term $\boldsymbol{N}=\boldsymbol{0}$ so the magnetic field is frozen into the plasma and the perpendicular plasma velocity $\boldsymbol{v}_{\perp}$ and magnetic field line velocities $\boldsymbol{w}_{\perp}$ are the same. 

Inside the diffusion region, $\boldsymbol{N}\neq \boldsymbol{0}$ and the magnetic field may slip through the plasma with the field line velocity defined as $$\boldsymbol{w}_{\perp} = \boldsymbol{v}_{\perp} + \frac{\boldsymbol{N}\times\boldsymbol{B}}{|\boldsymbol{B}|^2}\,.$$
In this paper, we do not derive the field line velocities in the diffusion regions of our models as this would require assumptions to be made about the non-ideal term $\boldsymbol{N}$ which has purposefully been left undefined such that the results of these models can be applied in a wide range of plasma regimes.

In the following sections the bifurcation models are revisited to consider whether reconnection on the separators is an integral part of the bifurcations by which the separators are formed.

\subsection{Null-null bifurcation separators (No pre-existing nulls)\label{sec:cluster-sep}}
The electric field for the first null-null bifurcation model is determined from Eqn~\ref{eqn:null-null-bifur} using Faraday's law  
\begin{eqnarray}
\label{eqn:null-null-1-E}
\boldsymbol{E} &=&  - \left(\frac12\delta^2 y\right) \;\hat{\boldsymbol{e}}_x + \left(\frac12\delta^2 x\right)\;\hat{\boldsymbol{e}}_y \nonumber \\
&& \qquad + \left(\frac12 (\gamma x^2 + \beta y^2) - \alpha xy\right)\;\hat{\boldsymbol{e}}_z\,,
\end{eqnarray}

Revisiting the null-null bifurcation examples described in Sec~\ref{sec:creation of nulls}, we note that in Example 1 where $a^2+bc>0$ a generic separator is always created at the same time as the two new improper 3D null points. In this example the separator lies between the two null points $(0,0,-\delta\sqrt{t})$ and $(0,0,\delta\sqrt{t})$, so it lies on the $z$-axis. The reconnection rate along this separator is therefore
\begin{eqnarray}
\int_{1sep} \frac{\boldsymbol{E}\cdot\boldsymbol{B}}{|\boldsymbol{B}|}\; {\rm d}l &=& \int_{-\delta \sqrt{t}}^{\delta \sqrt{t}} E_z(0,0,z)\, {\rm d}z \approx 0\,,
\end{eqnarray}
Thus, little to no reconnection is associated with the separator in this bifurcation model.

It is possible that a magnetic field containing a non-zero component of the electric field parallel to the separator can be found, but for a null-null bifurcation to occur no such electric field component is required.

\begin{figure}
  \centering
  \includegraphics[scale=0.35]{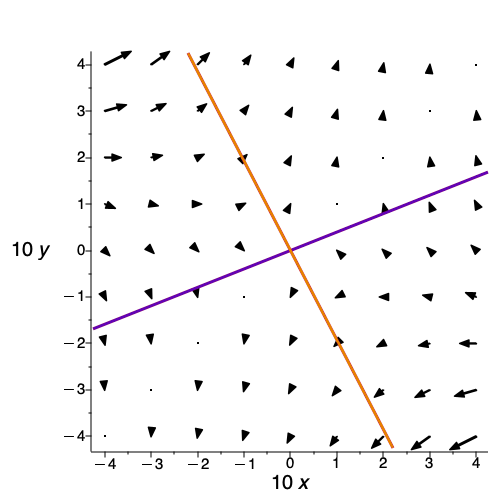}
  \caption{Cut in the $z=0$-plane for $-0.4 \leq x, y \leq 0.4$ at the instant of bifurcation, $t=0$, for the null-null bifurcation Example 1. The black arrows indicate the direction of the perpendicular plasma velocity in the plane and the purple and orange lines are where the separatrix surfaces cut this plane.}
\label{fig:null-null-rec1-vperp}
\end{figure}
From Eqn~\ref{eqn:vperp}, the plasma velocity perpendicular to the magnetic field may be determined. In the example considered above,
\begin{eqnarray*}
\boldsymbol{v}_{\perp} &=& \frac{ (px-qr) \;\hat{\boldsymbol{e}}_x + (py-qs) \;\hat{\boldsymbol{e}}_y -\delta^2(rx+ sy) \;\hat{\boldsymbol{e}}_z}{2|\boldsymbol{B}|}, \\
 \\
p &=&(z^{2}-\delta^{2}t) \delta^{2} \\
q &=& \beta y^{2}+\gamma x^{2}-2\alpha xy \\
r &=& c x-(a+z) y+t (-\alpha y+\gamma x) \\
s &=& (a-z) x+b y+t (\alpha  x-\beta  y)
\end{eqnarray*}
Figure~\ref{fig:null-null-rec1-vperp} shows a plot of the direction of $\boldsymbol{v}_\perp$ in the $z=0$ plane which is weak in strength close to the origin. However, the flow parallel to the $z=0$ plane is stagnation-like and would act to close the `X' formed by the intersection of the separatrix surfaces with this plane. Such a flow could potentially encourage the formation of of an electric field that would be important for reconnection.  

So, here, despite a stagnation type flow about the bifurcation point, the parallel electric field along the separator is zero. The limited reconnection associated with these short separators connecting a new pair of null points agrees with the rates of reconnection along such separators determined from analysis of numerical experiments\citep{Parnell2010ApJ} (see Figure~\ref{fig:flux-emerge-Epar}).

\subsection{Null-null bifurcation separators (Pre-existing nulls)\label{sec:cluster-sep-interclust}}
The electric field associated with the separators formed from a null-null bifurcation producing a new null pair in a magnetic field with existing null points can be determined from the model magnetic field presented in Eqn~\ref{eqn:null-null-bifur-intercluster}. The components of the electric field are
\begin{eqnarray*}
\boldsymbol{E}(\boldsymbol{x},t) &=& \frac{z_0 y}{2z_1}\left(4(bz-z_1^2)-(b^2-z_1^2) \right)\;\hat{\boldsymbol{e}}_x \\
&& +\frac{z_0 x}{2z_1}\left(-2(2b^2+z^2)-(b^2-z_1^2) \right)\;\hat{\boldsymbol{e}}_y \\
&& + \frac{2z_0xy}{z_1}\left(b-z\right)\;\hat{\boldsymbol{e}}_z\,.
\end{eqnarray*}

All the separators in this magnetic field lie along the $z$ axis which means that the electric field component relevant for determining the reconnection occurring along them is $E_z(0,0,z) \approx 0$. Thus, the reconnection rates along any of the separators created from null-null bifurcations are small.

\begin{figure}
  \centering
  \includegraphics[scale=0.35]{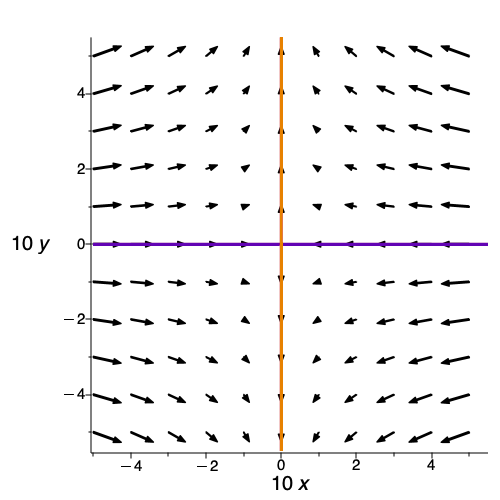}
  \caption{Cut in the $z=0$-plane for $-0.4 \leq x, y \leq 0.4$ at the instant of bifurcation, $t=0$, for the null-null bifurcation in a magnetic field with existing null points. The black arrows indicate the direction of the perpendicular plasma velocity in the plane and the purple and orange lines are where the separatrix surfaces cut this plane.}
\label{fig:null-null-rec2-vperp}
\end{figure}
The plasma velocity perpendicular to the magnetic field can be determined from Eqn~\ref{eqn:vperp}. The expressions for each component of this velocity are rather large so they are not presented here. However, from Figure~\ref{fig:null-null-rec2-vperp} it can be seen that $\boldsymbol{v}_\perp$ parallel to the $z=0$ plane is a stagnation type flow that runs parallel to the separatrix surfaces, as opposed to across them. Such a flow is not conducive to reconnection.

\subsection{Separator-separator bifurcation separators \label{sec:intercluster-sep}}
The time-dependent magnetic field (Eqn~\ref{eqn:sep-bifur-field}) used to illustrate a separator-separator bifurcation has a corresponding electric field of the form
\begin{eqnarray}
    \boldsymbol{E}(\boldsymbol{x},t) &=& \frac{xyz}{a^2}\frac{{\rm d}a}{{\rm d}t}\;\hat{\boldsymbol{e}}_x \nonumber \\
    && -4ca(t)(y^2+d^2)\left(\frac{x^3}{3}-a^2x\right)\frac{{\rm d}a}{{\rm d}t}\;\hat{\boldsymbol{e}}_y \nonumber \\
    && +2ay\frac{{\rm d}a}{{\rm d}t}\;\hat{\boldsymbol{e}}_z\,.
\end{eqnarray}
This electric field varies both spatially and temporally via $a(t)$. It also appears to be non-zero on the separator.

First, recall that $0<a(t)<0.5$ and observe that the separators curve a small distance away from, and then back to, the $y=0$ plane as they extend between the 3D null points located at $(-a(t),0,-b)$ and $(a(t),0,b)$.  The components of the electric field that contribute most to the integral of the parallel electric field along the separators are 
the $x$ and $z$ components. On the separators $x$ and $z$ both have the same sign with the separators crossing the $z=0$ plane on the $x=0$ axis (due to the symmetry in the model). Thus, we see that
$$\mathrm{sign}(E_z) = \mathrm{sign}(E_x) = \mathrm{sign}(y)\;,$$
since $a(t)>0$ and ${\rm d}a/{\rm d}t < 0\,.$

Since one separator lies in the $y \ge 0$ half of the domain and the other lies in the region where $y \le 0$, $E_{\|}$ will be directed in opposite directions along the separators since the magnetic field along both separators run from the positive null to the negative null. 

In particular, if we imagine the null points reside in distant clusters (e.g., we assume $b \gg a(t)$) then the $E_z$ component would become the dominant factor in determining $E_{\|}$ along the separators. Specifically, 
\begin{equation*}
   \int_{lsep, \; y\ge 0} \boldsymbol{E}_\|(l) dl  > 0 \quad {\mathrm{and}} \quad
   \int_{lsep, \; y\le 0} \boldsymbol{E}_\|(l) dl < 0\;.
\end{equation*}
The sign of $E_{\|}$ on a fieldline determines the direction of the reconnection, recovering the key finding from the numerical (resistive MHD) experiments\citep{Haynes2007,Parnell2010ApJ}, that the reconnection on the two new intercluster separators runs in opposite directions. Such electric fields would imply that reconnection at one separator transports flux into the new flux domain and the reconnection at the other transports it out. However, the model produces reconnection rates of the same magnitude along each separator, due to our over-simplified symmetric magnetic field. This is not what is found in the numerical experiments.

So, in contrast to a null-null bifurcation, a separator-separator bifurcation maybe associated with a significant amount of reconnection. This is apparent from the change of connectivity of the fieldlines forming the separatrix surfaces over time (Figure~\ref{fig:sep-sep bifurcation}). Before the bifurcation all the red fieldlines lying in the separatrix surface of the positive null exit through the top of the domain (Figure~\ref{fig:sep-sep bifurcation}a) but after the bifurcation a selection exit through the bottom and vice-versa for the blue fieldlines lying in the separatrix surface from the negative null (Figure~\ref{fig:sep-sep bifurcation}c). 

The bottom row of plots in Fig~\ref{fig:sep-sep bifurcation} show general fieldlines colour coded according to the topological domain in which they lie. Fieldlines that are contained within (lie above) the positive (red) separatrix surface are coloured pink: they enter and leave through the top of the box. Fieldlines that are contained within (lie below) the negative (blue) separatrix surface are coloured cyan: they enter and leave through the bottom of the box. Mustard-coloured fieldlines which represent the ambient field lie below the positive separatrix surface and above the negative separatrix surface (i.e. outside both surfaces) and enter through the bottom of the box and exit through the top. In the bottom left and middle plots of Figure~\ref{fig:sep-sep bifurcation} there are just these 3 types of fieldlines (excluding the special fieldlines that go through the null points). However, in the bottom right-hand plot of Figure~\ref{fig:sep-sep bifurcation} there are 4 types of fieldlines. A new type of fieldline (purple) lies above the positive (red) separatrix surface and below the negative (blue) separatrix surface: they enter through the top of the box and exit through the bottom. These have been formed by the reconnection of pink and cyan fieldlines at a separator. This reconnection creates a new domain containing these purple fieldlines and it also produces more mustard-coloured fieldlines.  

\begin{figure}
  \centering
  \includegraphics[scale=0.25]{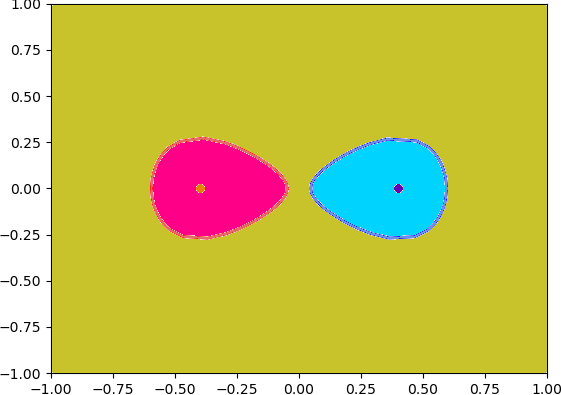}
   \includegraphics[scale=0.25]{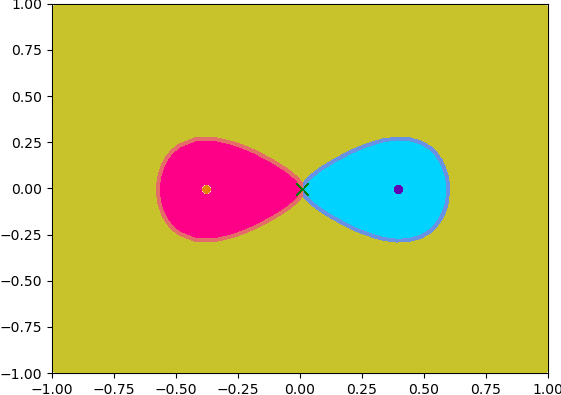}
   \includegraphics[scale=0.25]{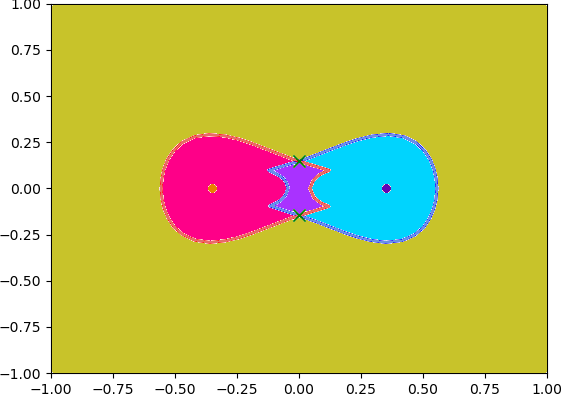}
  \includegraphics[scale=0.7]{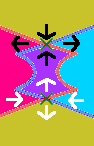}
  \caption{Cuts in the $z=0$-plane at $a(t)=0.4$, $a(t)=0.3874$ and $a(t)=0.35$ indicating the connectivity of fieldlines that thread this plane, where the colours used are the same as those used in Figure~\ref{fig:sep-sep bifurcation}. The lower right-hand panel shows a localised cut-out about the origin from the plot on the lower left (at $a(t)=0.35$ including the regions around the points where the separators thread this plane. Along with fieldline connectivity this plot shows the expected direction of the reconnection (arrows) at each separator.}
\label{fig:sep-rec-ill}
\end{figure}
Figure~\ref{fig:sep-rec-ill} shows a sketch of a cut through the magnetic field in the $z=0$ plane illustrating the flow of plasma across the separators. Note, all the fieldlines in the domain cross the $z=0$ plane: that is there are no fieldlines actually lying in the $z=0$ plane. From Figure~\ref{fig:sep-rec-ill} it can be seen that the reconnection about the separator in $y<0$ is such that magnetic field within the positive (red line) separatrix surface (shaded pink) reconnects with field from inside the negative (blue line) separatrix surface (shaded cyan). This results in a transfer of magnetic flux into both the newly formed domain that lies inside both the separatrix surfaces (shaded purple) and also the region that lies outside both separatrix surfaces (shaded mustard). About the other separator the transfer of flux is reversed. Naturally, for the new flux domain (purple) to form the rate of reconnection along the separator lying in the $y<0$ half of the domain must be greater than that along the other separator.

\begin{figure}
  \centering
  \includegraphics[scale=0.35]{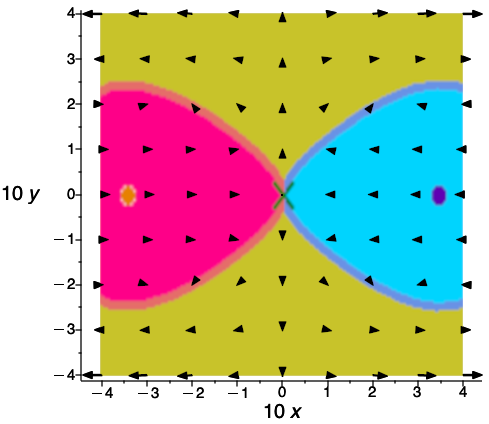}
  \caption{Cut in the $z=0$-plane for $-0.4 \leq x, y 0.4$ at the instant of bifurcation, $a(t)=0.3874$, where the black arrows indicate the direction resulting from the components of perpendicular plasma velocity $v_x$ and $v_y$ plotted over the top of the image showing the connectivity of fieldlines that thread this plane, coloured as in Figure~\ref{fig:sep-rec-ill}}
\label{fig:sep-rec-vperp}
\end{figure}
The plasma velocity $\boldsymbol{v}_\perp$ is also calculated using Eqn~\ref{eqn:vperp}. The equations are not presented here as they do not simplify easily, but since $\boldsymbol{E}$ and $\boldsymbol{B}$ have been given the components of $\boldsymbol{v}_\perp$ are straightforward to calculate. Figure~\ref{fig:sep-rec-vperp} shows a plot with arrows indicating the direction of $v_x$ and $v_y$ in the $z=0$ plane overlying the connectivity of fieldlines threading this plane at the instant of the bifurcation $a(t)=0.3874$. The arrows close to the origin form a stagnation type flow directed inwards to the origin along the $x$-axis and outwards along the $y$-axis. The bifurcation occurs at the origin in the $z=0$ plane. Here, though the separatrix surfaces that thread this plane form an `X' shape and, close to origin there is evidence of a flow perpendicular to these surfaces, pushing them together. This is the type of flow that acts to enhance both the electric field and electric current density around where the separatrix surfaces touch. Thus it is a flow conducive to reconnection.

\section{Formation Mechanisms of Null Clusters, Cluster and Intercluster Separators\label{sec:sep-form-mechanism}}

As seen in the numerical experiments, null clusters may have a zero or non-zero topological degree. Null clusters with zero topological degree have equal numbers of positive and negative null points. Since null points must be created and destroyed in pairs, the null points inside these null clusters are created in these weak-field regions via the null-null bifurcations. 

The null clusters with non-zero topological degree must have additional positive or negative nulls. For instance, null clusters A and D from Figure~\ref{fig:null-cluster-example} both have one negative null each, but A has two positive nulls and D has none leading to topological degrees for these null clusters of -1 and 1, respectively. In order for null clusters like this to form a null-null bifurcation must occur, creating a positive and a negative null point. Then, due to plasma flows, the two null points may be moved apart such that the original null cluster with two null points is split into two separate null clusters. Note, that null clusters can also merge together creating a new cluster with a topological degree equal to the sum of topological degrees of the two original null clusters.

Inside null clusters, null points are typically found either to be connected together by separators like beads on a string (as seen in Figure~\ref{fig:flux-emerge}), or they may not be connected to any other null points. So far, from numerical experiments, any pair of opposite polarity null points inside a null cluster have only been found to be connected by a single separator. This suggests that cluster separators are formed via null-null bifurcations in which single separators are formed. Such separators all form at the appearance of a new null pair with a separator (i) connecting the newly created null points and/or (ii) connecting a newly emerged null point to a pre-existing null point. This leads to the ``beads on a string'' nature of null points within a null cluster. 

Due to the absence of any two null points within the same null cluster observed to be connected by multiple separators it is unlikely that many/any cluster separators are formed by separator-separator bifurcations. This is likely to be due to the lack of significant/appropriate flows within a null cluster which push the separatrix surfaces from two existing null points together causing a separator-separator bifurcation.

Intercluster separators that connect null points from different null clusters can arise through several mechanisms. The mechanism that creates single intercluster separators is a null-null bifurcation occurring in a (or creating a new) null cluster leading to the creation of a separator connecting one of the new nulls to a null point in a different pre-existing null cluster. This mechanism created the first intercluster separator seen in the flux emergence experiment\cite{Parnell2010ApJ} and also the mechanism that created the single interclusters connecting nulls between the null clusters A \& C, B \& C and B\& D in the magnetosphere experiment\cite{Haynes2010}. 

The single intercluster separator between clusters A \& D was already evident in the initial frame of the experiment. In theory it could have arisen in the same way as the other single intercluster separators or as a short separator connecting a new null pair that is then stretched splitting up into the two null clusters A and D. 

Finally, the double separator connecting null clusters A \& C in the magnetosphere experiment is created by the separator-separator bifurcation as are most of the intercluster separators found in the flux emergence experiment.

\section{Conclusions\label{sec:conclusions}}
A necessary and sufficient condition for 3D magnetic reconnection is that there exists a region in which the electric field parallel to the magnetic fieldlines is non-negligible\citep{Schindler88,Hesse88}. In a few numerical resistive MHD experiments the 3D magnetic reconnection sites have been located by determining the integral of the parallel electric field along fieldlines, as well as the topological features (e.g. null points, separatrix surfaces, spines and separators). These experiments reveal that the fieldlines with the largest integral of parallel electric field are separators. However, there are some separators, usually those short in length, that have a small integral of parallel electric field suggesting they are not associated with much reconnection. 

Further analysis of the numerical experiments\citep{Parnell2010ApJ,Haynes2010} revealed what appeared to be two different types of separators: cluster separators that link positive and negative null points that reside within a single weak field region (a null cluster); and intercluster separators that connect null points of opposite sign from two distinct weak field regions. The latter are much longer than the former and are typically associated with significant amounts of magnetic reconnection. By considering kinematic models of the bifurcations that lead to the creation of these two types of separator it has been possible to explain their key characteristics. The main properties of these two types of separators and their reconnection are presented in Table~\ref{tab:sep-properties}.

Cluster separators are created within a single weak magnetic field region via a null-null bifurcation that also forms a pair of opposite-polarity null points. Single separators are formed during this type of bifurcation that may connect the two newly formed null points and can also connect a newly formed null point to a pre-existing null point within the same null cluster. All of these separators are short in length as they are contained with a single null cluster. The separators connecting the newly created null points start out infinitesimally short and grow in length as the new null points move apart.
Although at the instant that the new nulls are created reconnection occurs, once they are formed little reconnection seems to occur about these types of separator. 

\begin{table}
\caption{\label{tab:sep-properties} Properties of Separators.}
\begin{ruledtabular}
    \begin{tabular}{ccc}
        Property & Cluster Separators & Intercluster Separators \\
        \hline
        Location & \textit{Inside a single} & \textit{Connect different} \\
        & \textit{null cluster} & \textit{null clusters}  \\
        Typically created & Singularly & Singularly or in pairs \\
        Bifurcation & Null-null & Null-null or Sep-sep \\
    %    Null behaviour & Within a null cluster any two oppositely signed nulls are typically connected by a single separator & Between null clusters, any two null points may be connected by a large number of separators \\
        Separator significance & Locally & Globally \\
        Length & Short  & Long \\
        $\boldsymbol{E}_\|$ & Weak  & Often strong \\
        $\int |\boldsymbol{E}_\| | \mathrm{d}s$ & Small & Large \\
        Impact of reconnection & Little & Globally significant
       \end{tabular}
\end{ruledtabular}
\end{table}
The kinematic models presented here show that the perpendicular plasma velocity is a stagnation flow about the separator which may, in some circumstances, act to close the `X' formed by the intersection of the separatrix surfaces. Such a flow could lead to the creation of conditions suitable for reconnection about the separator.

In Section~\ref{sec:separator reconnection}, it was noted that  $\boldsymbol{E}\approx -\partial \boldsymbol{A}/\partial t$ provided that the length scales are much longer than the Debye length. At the moment of bifurcation, and for a short time afterwards, this assumption fails. This means that the reconnection rate immediately after the bifurcation along the separator connecting the two null points is equal to the change in electric potential $\phi$ along this separator. Once the separators are much longer than a Debye length, little reconnection would be expected due to the null-null bifurcation. However, flows in the system not associated with the bifurcation may build up currents at the separators and drive reconnection there. Within a null cluster, where the separators are short, any driven reconnection is likely to be weak.

Intercluster separators are different in that they connect null points that lie in different (distinct and typically distant) null clusters. These separators are either formed in pairs via separator-separator bifurcations or singularly via a null-null bifurcation where one of the newly created null points forms a separator connecting it to a pre-existing null. The singularly-formed intercluster separators are not associated with much reconnection immediately after their formation, as discussed above but, since these separators extend over large distances, flows elsewhere in the system may lead to the build up of significant currents at these separators leading to significant reconnection.

The separators created in pairs arise from a bifurcation which has been shown in the kinematic model presented here to be associated with a perpendicular plasma velocity that would drive the separatrix surfaces together enabling the build up of an electric field. From the kinematic model the integrated parallel electric field along the separators shows they would be associated with a significant amount of reconnection, at least shortly after they have been formed.      
After the bifurcation, it is seen that the reconnection continues on these separators and leads to the creation of a new, topologically distinct flux domain. This new flux domain grows in size as the flux in it increases. Further reconnection may be driven by external flows since these separators typically extend over significant distances.

The separator-separator kinematic model considered here reveals that the electric current parallel to the separators is directed up one separator and down the other, thus forming a closed ring. A consequence of this is that the resulting magnetic reconnection on one separator acts to fill the newly created flux domain, whilst on the other separator the direction is reversed so it acts to empty the new flux domain in agreement with what has been found through analysis of numerical experiments\citep{Haynes2007,Parnell2008,Parnell2010ApJ,Parnell2010JGR,Wilmot-Smith2011}. If the rates of reconnection are the same along these two separators then, at any instant, the same amount flux will be transported both into and out of the newly formed domain bounded by the two separators and so this domain will not grow in size. Hence, the reconnection rates must be different along the two separators in order to increase or decrease the flux in the domain enclosed by the pair of separators. Due to symmetry the kinematic model presented here produces the same reconnection rates along the two separators.

These results highlight the importance of, at least some, intercluster separators as sites of significant reconnection. Given that there have so far been few studies of this process, there is clearly a need to carry out more self-consistent numerical experiments to enhance our understanding of it. Specifically, it would be interesting to see whether the singularly formed intercluster separators are the ones seen in numerical experiments that are associated with less reconnection than the ones created in pairs. This may not be the case as, after formation, the reconnection occurring at these separators may be driven by global plasma flows. Indeed, since the separators created in pairs have different rates of reconnection along them it may be just this difference that leads to the intercluster separators with considerably different reconnection rates. 

\begin{acknowledgments}
The author would like to thank both referees, as well as Profs T.\ Neukirch \& E.R.\ Priest for useful suggestions and questioning of the results which led to important improvements of this paper.
\end{acknowledgments}

\end{thebibliography}
%\bibliography{3DSepRecon1}

\end{document}